%% file: arXiv_v3.tex
\documentclass[peerreview]{IEEEtran}

\makeatletter
\let\warning\@latex@warning
\makeatother

\input{preambles}

\usepackage{hyperref} 
\hypersetup{
    colorlinks,
    linkcolor={blue!80!black},
    citecolor={green!50!black},
    urlcolor={blue!80!black}
}

\title{Key Assistance, Key Agreement, and Layered Secrecy for Bosonic Broadcast Channels}

\author{
		\vspace{0.1cm}
    \IEEEauthorblockN{Uzi Pereg\IEEEauthorrefmark{1}, 
		 Roberto Ferrara\IEEEauthorrefmark{1}, and Matthieu R. Bloch\IEEEauthorrefmark{2}} \\
\IEEEauthorrefmark{1}Institute for Communications Engineering, Technical University of Munich \\
     \IEEEauthorrefmark{2}School of Electrical and Computer Engineering, Georgia Institute of Technology\\
     Email: {\tt $\{$uzi.pereg,roberto.ferrara$\}$@tum.de, matthieu.bloch@ece.gatech.edu}
}

\begin{document}
\maketitle

\begin{abstract} 
Secret-sharing building blocks based on quantum broadcast communication are studied.
The confidential capacity region of the pure-loss bosonic broadcast channel is determined, both with and without key assistance, 
and  an achievable region is established for the lossy bosonic broadcast channel.
If the main receiver has a transmissivity of $\eta<\frac{1}{2}$,  then confidentiality solely relies on the key-assisted encryption of the one-time pad.
We also address conference key agreement for the distillation of two keys, a public key and a secret key.
A regularized  formula is derived for the key-agreement capacity region in finite dimensions. In the %
bosonic case, the key-agreement region is included within the capacity region of the corresponding broadcast channel with confidential messages. We then consider a network with layered secrecy, where three users with different security ranks communicate over the same broadcast network.
We derive an achievable layered-secrecy region for a pure-loss bosonic channel that is formed by the concatenation of two beam splitters.
\end{abstract}


\section{Introduction}
\setcounter{page}{1} 
Physical-layer security requires the communication of private information to be secret regardless of the computational capabilities of a potential eavesdropper \cite{RezkiZorguiAlomairAlouini:17p,BlochBarros:11b}.  Secret-key agreement is a promising method to achieve this goal, whereby the sender and the receiver generate a secret key before communication takes place. 
Maurer \cite{Maurer:93p} and Ahlswede and Csisz\'ar \cite{AhlswedeCsiszar:93p1} have independently developed and analyzed 
the information-theoretic model for such a protocol, whereby Alice and Bob use pre-existing correlations, along with a public insecure channel, to generate a secret key.  
Devetak and Winter \cite{DevetakWinter:05p} considered the quantum counterpart of key distillation from a shared quantum state and public classical communication.
Conference key agreement protocols \cite{MurtaGrasselliKampermannBruss:20p}, also known as multi-party key distribution are particularly relevant to this work.
In this framework, the aim is to distribute a common key between several users, allowing them to broadcast secure messages in a network (see also \cite{Berkovits:91c,Tzeng:02p}).
In practice, quantum key distribution (QKD) is among the most mature quantum technologies with an information-theoretic basis  \cite{BennettBrassard:14p}, as it is already implemented in experiments 
\cite{JKLGD:13p,WYCHSLZZGH:15p,
PKBJSAAMCH:17p,LWWFLLLZZL:19l},
and in commercial use as well \cite{Qui:14p,DiamaniLoQiYuan:16p}.
Most QKD implementations are based on optical communication, either with  optical fibers or in free space \cite{ZXCPP:18p,BedingtonArrazolaLing:17p}.
A QKD protocol aims to distribute a secret symmetric key between authorized partners, %
with no assumption regarding the communication channel but the laws of quantum mechanics \cite{SBCDLP:09p,PABBCEGLO:20p}.  The key can later be used to communicate   using classical encryption schemes, %
such as the one-time pad (OTP) cypher. %
As shown by Shannon \cite{Shannon:49p1},
information-theoretic security can be guaranteed if and only if the entropy of the key string is at least as large as  the message length. 
Usually, the cryptographic analysis is not restricted to a given noise model.
Here, on the other hand, we will incorporate the OTP cypher within our network coding scheme for communication over a noisy channel. %
Classical channel coding with key assistance, \ie given a pre-shared key, is studied, \eg in 
\cite{Yamamoto:97p,KangLiu:10c,SchaeferKhistiPoor:18p}.

In some noise models, assuming that the channel statistics are known, communication can also be secured without key assistance.
The broadcast channel with confidential messages is a network setting that involves transmission of information to two users, such that part of the information should be accessible for both users, while the other part is only intended for one of them. In commercial terms, those components can be thought of as basic and premium packages, where the latter may require an additional subscription fee. Confidentiality requires that the non-subscribed receiver  cannot decode the private component.
In the classical model,
the sender transmits a sequence $X^n$ over a given memoryless broadcast channel $p_{Y,Z|X}$, such that the output sequences $Y^n$ and $Z^n$ are decoded by two independent receivers. The transmission encodes two types of messages, a common message   sent to both receivers at rate $R_0$, and a private message   sent to Receiver $Y$ at rate $R_1$, while  eavesdropped by Receiver $Z$.
   This model was first introduced by Csisz\'ar and K{\"o}rner \cite{CsiszarKorner:78p}, who showed that the confidential capacity  region is given by
\begin{align}
\opC(p_{Y,Z|X})=%
\bigcup_{ p_{U,V} p_{X|V} } \left\{
\begin{array}{rl}
(R_0,R_1)\,:\; R_0\leq& \min\left( I(U;Y)\,,\; I(U;Z) \right)\\
							 R_1\leq& I(V;Y|U)-I(V;Z|U)
\end{array} \right\}
\label{eq:Clbc}
\end{align}
where $U$ and $V$ are auxiliary random variables. A recent overview of information-theoretic security and its applications can be found in
\cite{BlochGunluYener:21p}
(see also \cite{LiangPoorShamai:09n,SchaeferBocheKhistiPoor:17b}). The broadcast channel with layered decoding and secrecy is a generalization of the degraded broadcast channel with confidential messages \cite{LyLiuBlankenship:12p,ZouLiangLaiShamai:15p1,ZouLiangLaiPoorShamai:15p}. 
The model describes a network in which multiple users have different credentials to access confidential information. 
Zou \etal \cite{ZouLiangLaiPoorShamai:15p} give two examples for practical applications. The first example is a  WiFi network of an agency, in which a user is allowed to receive files up to a certain security clearance but should be kept ignorant of classified files that require a higher security level. 
As pointed out in \cite{ZouLiangLaiPoorShamai:15p}, the agency can set the channel quality on a clearance basis by assigning more communication resources
to users with a higher security clearance. 
The second example in \cite{ZouLiangLaiPoorShamai:15p} is a  social network in which one user wishes to share more information with close friends and less information with others. 
Tahmasbi \etal \cite{TahmasbiBlochYener:20p} have further demonstrated that, for some adversarial models with feedback, the layered-secrecy structure 
 allows the provision of secrecy in hindsight, deferring 
the decisions as to which bits are secret to a later stage.
Connections between confidential channel codes and other cryptographic protocols, such as advantage distillation, information reconciliation, and privacy amplification, can be found \eg in 
\cite{WieseBoche:21a,BlochBarrosRodriguesMcLaughlin:08p,BlochBarros:11b}.

\vspace{-0.75cm}
\begin{center}
\begin{figure}[b]
\includegraphics[scale=0.85,trim={-5cm 11.5cm 9cm 11.5cm},clip]
{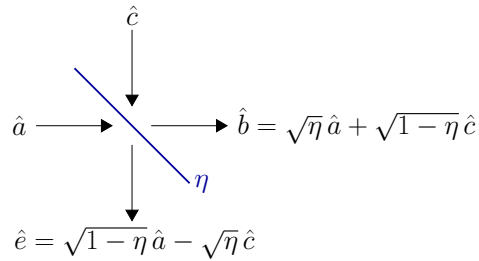} %
\caption{The beam splitter relation of the single-mode bosonic broadcast channel.
}
\label{fig:BSp}
\end{figure}
\end{center}

Optical communication forms the backbone of the Internet \cite{BardhanShapiro:16p,Pereg:21c2,Savov:12z,
KumarDeen:14b}.
The bosonic channel is a simple quantum-mechanical model for optical communication over free space or optical fibers 
\cite{WPGCRSL:12p,WildeHaydenGuha:12p}.
 An optical communication system consists of a modulated source of photons, the optical channel, and an optical detector. %
For a single-mode bosonic broadcast channel, the channel input is an electromagnetic field mode with  annihilation operator $\ha$, and the output is a pair of modes with  annihilation operators $\hb$ and $\he$.
The annihilation operators  %
correspond to the transmitter (Alice), the legitimate receiver of the common and confidential information (Bob), and the receiver that eavedrops on the confidential information (Eve), respectively.
The input-output relation of the bosonic broadcast channel in the Heisenberg picture \cite{HolevoWerner:01p} is given by %
\begin{align}
\hb&=\sqrt{\eta}\, \ha +\sqrt{1-\eta}\,\hc \\
\he&=\sqrt{1-\eta}\, \ha -\sqrt{\eta}\,\hc
\end{align}
where $\hc$ is associated with the environment noise and the parameter $\eta$  is %
the transmissivity, $0\leq \eta\leq 1$, which captures, for instance, the length of the optical fiber and its absorption length \cite{EisertWolf:05c}. The relation above corresponds to the outputs of a beam splitter, as illustrated in Figure~\ref{fig:BSp}.
The bosonic channel can be viewed as the quantum counterpart of the classical channel with additive white Gaussian noise (AWGN),
which is a well-known model in classical communications \cite{BCCGPP:07b}. Among others, the Gaussian broadcast channel describes the wide-band thermal noise in the receiver electronic circuits for two remote receiving antennas \cite{Kogan:96b}.
As the bosonic broadcast channel, from $A$ to $BE$ (jointly), is isometric,  it does not model the distortion introduced by the communication medium \cite{Shapiro:09p}. 
Instead, the bosonic broadcast channel models the de-modulation process at the destination location, where the optical signal is converted into two signals for two independent users by a beam splitter. In a lossy bosonic channel, the noise mode $\hc$ is in a Gibbs thermal state, while, in a pure-loss bosonic channel, the noise mode is in the vacuum state. The channel is called `lossy' or `pure-loss' since the marginal channels, from $A$ to $B$, and from $A$ to $E$, are non-reversible and involve loss of photons in favor of the other receiver.

The broadcast channel with confidential messages can be viewed as a generalization of the wiretap channel 
\cite{CsiszarKorner:78p} \cite[Section 22.1.3]{ElGamalKim:11b}.
 Devetak \cite{Devetak:05p} and Cai \etal \cite{CaiWinterYeung:04p} addressed the quantum wiretap channel without key assistance and established a regularized characterization of the secrecy capacity. Connections to the coherent information of %
a quantum point to point channel were drawn in \cite{DevetakWinter:05p}. In general, the secrecy capacity is not additive, hence regularization is necessary \cite{LiWinterZouGuo:09,ElkoussStrelchuk:15p}.
A single-letter characterization was established in the special cases of entanglement-breaking channels \cite{WildeHsieh:12p}, as well as the less noisy and more capable wiretap channels \cite{Watanabe:12p}. 
 Qi \etal \cite{QiSharmaWilde:18p} determined the entanglement-assisted secrecy capacity of the quantum wiretap channel
(see also \cite{HsiehWilde:10p,SharmaWakauwaWilde:17a}). 
 Davis \etal \cite{DavisShirokov:18p} considered an energy-constrained setting, and  Boche \etal 
\cite{BocheCaiNotzelDeppe:19p,BochCaiDeppeNotzel:17p} studied the quantum wiretap channel with an active jammer.
Hsieh \etal \cite{HsiehLuoBrun:08p} and Wilde \cite{Wilde:11p} presented a regularized formula for the secret-key-assisted quantum wiretap channel. %
Furthermore, the capacity-equivocation region was established, characterizing the tradeoff between secret key consumption and private classical communication \cite{HsiehLuoBrun:08p,Wilde:11p} (see also \cite{WildeHsieh:12p}\cite[Section 23.5.3]{Wilde:17b}).
In \cite{Devetak:05p}, Devetak considered entanglement generation using a secret-key-assisted quantum channel. The quantum 
Gel'fand-Pinsker wiretap channel is considered in \cite{AnshuHayashiWarsi:18c} and
other related scenarios can be found in \cite{KonigRennerBariskaMaurer:07p,GHKLLSTW:14p,LupoWildeLloyd:16p}. Secrecy in the form of quantum state masking was recently considered in \cite{PeregDeppeBoche:21p}.
 Key distillation is further considered  in 
\cite{TakeokaSeshadreesanWilde:17p,PirandolaGarciaBransteinLloyd:09p,PLOB:17p}.
Quantum broadcast %
channels were studied in various settings as well, \eg
 \cite{YardHaydenDevetak:11p,SavovWilde:15p,
WangDasWilde:17p,DupuisHaydenLi:10p,%
BaumlAzuma:17p,%
BochCaiDeppe:15p,Hirche:12z,XieWangDuan:18c,Palma:19p,AnshuJainWarsi:19p1,ChengDattaRouze:19a}. %
Yard \etal \cite{YardHaydenDevetak:11p} derived the superposition inner bound and determined the capacity region for the degraded classical-quantum  broadcast channel. Entanglement generation is considered in \cite{YardHaydenDevetak:11p} as well. %
Wang \etal \cite{WangDasWilde:17p} used the previous characterization to determine the capacity region for Hadamard broadcast channels as well.
Dupuis \etal \cite{DupuisHaydenLi:10p} developed the entanglement-assisted version of Marton's region for users with independent messages.
Bosonic broadcast channels are considered in \cite{GuhaShapiro:07c,GuhaShapiroErkmen:07p,DePalmaMariGiovannetti:14p,TakeokaSeshadreesanWilde:16c,TakeokaSeshadreesanWilde:17p,LaurenzaPirandola:17p,AndersonGuhaBash:21c}. The quantum broadcast and multiple access channels with confidential messages were recently considered by Salek \etal \cite{SalekHsiehFonollosa:19a,SalekHsieFonollosa:19c} and 
Aghaee \etal \cite{AghaeeAkhbari:19c}, respectively (see also \cite{BochJanssenSaeedianaeeni:19a}). Other security settings of transmission over bosonic channels include covert communication \cite{BGPHGTG:15p,BullockGagatsosGuhaBash:20c,BullokGagatsosGuhaBash:20p}, optical QKD \cite{
PLOB:17p,Pirandola:19p,Pirandola:19p1,TahmasbiBloch:20p}, and entanglement distillation \cite{ZhangVanLoock:20p,TakeokaSeshadreesanWilde:16c}.

In this paper, we study secrecy-sharing building blocks that are based on quantum broadcast communication. We begin with the quantum broadcast channel  with confidential messages. We consider two scenarios, either with or without shared key assistance.
In particular, we determine the confidential capacity region of the pure-loss bosonic broadcast channel in both settings, as depicted in Figure~\ref{fig:BosonicKey}, under the assumption of the long-standing minimum output-entropy conjecture, and we establish an achievable region for the lossy bosonic channel. The main technical challenge is in the converse proof, which requires the conjecture.
The achievability proof is based on rate-splitting, combining the ``superposition coding" strategy with the OTP cypher using the shared key.
The converse proof relies on the long-standing minimum output-entropy conjecture, which is known to hold in special cases \cite{Palma:19p}.
Without key assistance, confidential transmission is only possible if Bob's channel has a higher transmissivity than Eve's channel, \ie $\eta>\frac{1}{2}$. 
Otherwise, if Bob's channel is noisier than Eve's, \ie $\eta<\frac{1}{2}$, then confidentiality solely relies on the key-assisted encryption of the OTP.

Next, we address key agreement for the distillation and distribution of two keys. The public key is distributed between Alice, Bob, and Eve, while the confidential key is only meant for Alice and Bob, and  must be hidden from Eve.
Such a protocol can be viewed as a  conference key agreement  \cite{MurtaGrasselliKampermannBruss:20p}, or multi-party key distribution. 
We obtain a regularized formula for the key-agreement capacity region for the distillation of public and secret keys.
We then consider quantum layered secrecy, whereby Alice communicates with three receivers, Bob, Eve 1, and Eve 2. The information has different  confidentiality layers, which are labeled by `0', `1', and '2'. 
In Layer 0, we have a common message that is intended for all three receivers.
In the next layer, the confidential message is decoded by Bob and Eve 1 but should remain hidden from Eve 2.
Finally, the top-secret message of Layer 2 is only decoded by Bob, while remaining confidential from both Eve 1 and  Eve 2. 
 We derive a regularized formula for the layered-secrecy capacity region of the degraded quantum broadcast channel in finite dimensions and an achievable region for the pure-loss bosonic broadcast channel.

\section{Definitions and Related Work}
\label{sec:def}
\subsection{Notation, States, and Information Measures}
\label{subsec:notation}
 We use the following notation conventions. %
Script letters $\Xset,\Yset,\Zset,...$ are used for finite sets.
Lowercase letters $x,y,z,\ldots$  represent constants and values of classical random variables, and uppercase letters $X,Y,Z,\ldots$ represent classical random variables.  
 The distribution of a  random variable $X$ is specified by a probability mass function (pmf) 
	$p_X(x)$ over a finite set $\Xset$. %
 We use $x^j=(x_1,x_{2},\ldots,x_j)$ to denote  a sequence of letters from $\Xset$. %
 A random sequence $X^n$ and its distribution $p_{X^n}(x^n)$ are defined accordingly. 
The type $\hP_{x^n}$ of a given sequence $x^n$ is defined as the empirical distribution $\hP_{x^n}(a)=N(a|x^n)/n$ for $a\in\Xset$, where $N(a|x^n)$ is the number of occurrences of the symbol $a$ in the sequence $x^n$.
A type class is denoted by $\tau_n(\hP)=\{ x^n \,:\; \hP_{x^n}=\hP \}$.
For a pair of integers $i$ and $j$, $1\leq i\leq j$, we write a discrete interval as $[i:j]=\{i,i+1,\ldots,j \}$. %
In the continuous case, we use the cumulative distribution function 	$F_Z(z)=\prob{Z\leq z}$ for $z\in\mathbb{R}$, or alternatively, the probability density function (pdf) $f_Z(z)$,  when it exists. 
We write $Z\sim\mathcal{N}_{\mathbb{R}}(\mu,\sigma^2)$ to indicate that $Z$ is 
a real-valued Gaussian variable, with %
$f_Z(z)=\frac{1}{\sqrt{2\pi\sigma^2}}e^{-(z-\mu)^2/2\sigma^2}$.
  A complex-valued Gaussian variable $\alpha\sim\mathcal{N}_{\mathbb{C}}(0,\sigma^2)$ can be expressed as $\alpha=X+\textrm{i}Y$ where 
	$X,Y\sim\mathcal{N}_{\mathbb{R}}(0,\sigma^2)$ are statistically independent. 

The state of a quantum system $A$ is a density operator $\rho$ on the Hilbert space $\Hset_A$.
A density operator is an Hermitian, positive semidefinite operator, with unit trace, \ie 
 $\rho^\dagger=\rho$, $\rho\succeq 0$, and $\trace(\rho)=1$.
The trace distance between two density operators $\rho$ and $\sigma$ is $\norm{\rho-\sigma}_1$ where $\norm{F}_1=\trace(\sqrt{F^\dagger F})$.
Define the quantum entropy of the density operator $\rho$ as
$%
H(\rho) \triangleq -\trace[ \rho\log(\rho) ]
$.
Consider the state of a pair of systems $A$ and $B$ on the tensor product $\Hset_A\otimes \Hset_B$ of the corresponding Hilbert spaces.
Given a bipartite state $\sigma_{AB}$, %
define the quantum mutual information as
\begin{align}
I(A;B)_\sigma=H(\sigma_A)+H(\sigma_B)-H(\sigma_{AB}) \,. %
\end{align} 
Furthermore, conditional quantum entropy and mutual information are defined by
$H(A|B)_{\sigma}=H(\sigma_{AB})-H(\sigma_B)$ and
$I(A;B|C)_{\sigma}=H(A|C)_\sigma+H(B|C)_\sigma-H(A,B|C)_\sigma$, respectively.

A detailed description of (continuous-variable) bosonic systems can be found in \cite{WPGCRSL:12p}. Here, we only define the notation for the quantities that we use.
We use hat-notation, \eg $\ha$, $\hb$, $\he$, to denote operators that act on a quantum state.
The single-mode Hilbert space is spanned by the Fock basis 
$\{ \ket{n} \}_{n=0}^\infty$. Each $\ket{n}$ is an eigenstate of the number operator $\hn=\ha^\dagger \ha$, where $\ha$ is the bosonic field annihilation operator. In particular,  $
\ket{0}$ is the vacuum state of the field. 
The \emph{creation operator} $\ha^{\dagger}$ creates an excitation: 
$\ha^{\dagger}|n\rangle=\sqrt{n+1}|n+1\rangle$, for $n\geq 0$. Reversely, the \emph{annihilation operator} $\ha$ takes away an excitation: $\ha|n+1\rangle=\sqrt{n+1}|n\rangle$. 
A coherent state $|\alpha\rangle$, where $\alpha\in\mathbb{C}$,  
corresponds to an oscillation of the electromagnetic field, and it is the outcome of applying the displacement operator to the vacuum state, \ie $|\alpha\rangle=D(\alpha)|0\rangle$, 
which resembles the creation operation, with $D(\alpha)\equiv \exp(\alpha \ha^\dagger-\alpha^* \ha)$. 
 A thermal state $\tau(N)$ is a Gaussian mixture of coherent states,  where
\begin{align}
\tau(N)\equiv \int_{\mathbb{C}} d^2 \alpha \frac{e^{-|\alpha|^2/N}}{\pi N} \ketbra{\alpha} = \frac{1}{N+1}\sum_{n=0}^{\infty} \left(\frac{N}{N+1}\right)^n \ketbra{n}
\label{eq:tau}
\end{align}
given an average photon number $N\geq 0$.

\subsection{Quantum Broadcast Channel}
\label{subsec:Qchannel}
A quantum broadcast channel maps a quantum state at the sender system to a quantum state at the receiver systems. 
Here, we consider a channel with two receivers.
Formally, a quantum broadcast channel  is a   linear, completely positive, trace-preserving map 
$%
\channel_{ A\rightarrow B E}  %
$ %
corresponding to a quantum physical evolution.
We assume that the channel is memoryless. That is, if the system $A^n=(A_1,\ldots,A_n)$ are sent through $n$ channel uses, then the input state $\rho_{ A^n}$ undergoes the tensor product mapping
$%
\channel_{ A^n\rightarrow B^n E^n}\equiv  \channel_{ A\rightarrow B E}^{\otimes n} %
$. %
The marginal channel $\channel^{(1)}_{A\rightarrow B}$ is defined by
\begin{align}
\channel_{A\rightarrow B}^{(1)}(\rho_A)=\trace_{E} \left( \channel_{ A\rightarrow B E}(\rho_{A}) \right) 
\end{align}
for Receiver 1, and similarly $\channel_{A\rightarrow E}^{(2)}$ for Receiver 2. 
One may say that $\channel_{A\rightarrow B E}$ is an extension of $\channel^{(1)}_{A\rightarrow B}$ and $\channel^{(2)}_{A\rightarrow E}$.
 The transmitter, Receiver 1, and Receiver 2 are often referred to as Alice, Bob, and Eve, respectively.

A quantum broadcast channel is \emph{degraded}\begin{footnote}{This definition generalizes the classical notion of a stochastically degraded broadcast channel.}\end{footnote} if there exists a degrading channel 
$\Dset_{B\rightarrow E}$ such that 
\begin{align}
    \channel^{(2)}_{A\rightarrow E}\equiv \Dset_{B\rightarrow E} \circ \channel^{(1)}_{A\rightarrow B} \,.
\end{align}
We also say that Eve's channel $\channel^{(2)}_{A\rightarrow E}$ is degraded with respect to Bob's channel $\channel^{(1)}_{A\rightarrow B}$.
Intuitively, this means that Eve receives a noisier signal than Bob.
In the opposite direction, a broadcast channel is called \emph{reversely degraded} if $\channel^{(1)}_{A\rightarrow B}$ is degraded with respect to $\channel^{(2)}_{A\rightarrow E}$.

\begin{center}
\begin{figure}[tb]
\includegraphics[scale=0.85,trim={-1cm 11.5cm 2cm 11.5cm},clip]
{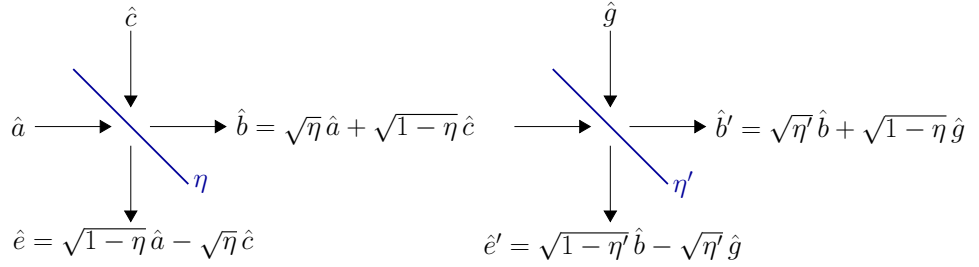} %
\caption{The degraded representation of the bosonic broadcast channel.
Bob's output mode is directed into another beam splitter with transmissivity $\eta'=\frac{1-\eta}{\eta}$. 
The state of the output mode $\he'$ is the same as that of $\he$.
}
\label{fig:BSpDegraded}
\end{figure}
\end{center}

Consider a quantum channel $\mathcal{N}_{A\rightarrow B}$ with a single receiver.
Every quantum channel $\mathcal{N}_{A\rightarrow B}$ has 
a Stinespring dilation $\Uset^{\,\mathcal{N}}_{ A\rightarrow B K}$, where $K$ is a reference system which
 is often associated with the receiver's environment. 
The broadcast channel $\Uset^{\,\mathcal{N}}_{ A\rightarrow B K}$ is an isometric extension of $\mathcal{N}_{A\rightarrow B}$, \ie
\begin{align}
&\Uset^{\,\mathcal{N}}_{A\rightarrow BK}(\rho_{A})=U\rho_{A} U^\dagger \\
& {\mathcal{N}}_{ A\rightarrow B }(\rho_{A})=\trace_{K}( U\rho_{A} U^\dagger )
\end{align}
where the operator $U$ is an isometry, \ie $ U^\dagger U=\identity$.
The channel $\widehat{\mathcal{N}}_{A\rightarrow K}(\rho_{A})=\trace_B( U\rho_{A} U^\dagger )$ is called a complementary channel for
${\mathcal{N}}_{ A\rightarrow B}$. 
\vspace{-0.1cm}
\begin{remark}
\label{rem:bosonicDeg}
If the broadcast channel $\channel_{A\rightarrow BE}$ is isometric, \ie $\channel_{A\rightarrow BE}(\rho_A)=U\rho_A U^\dagger$ for some isometry $U$, then the marginal channels are complementary to each other. In this case Eve's system $E$ can be thought of as Bob's environment. In particular, this is the case for the bosonic broadcast channel. The bosonic isometry is specified by  \cite{Palma:19p}
\begin{align}
   U_\eta=\exp\left( (\ha^\dagger \hc-\hc^\dagger \ha)\arccos\sqrt{\eta} \right) 
\,.
\end{align}
The bosonic broadcast channel is degraded if $\eta\geq \frac{1}{2}$, and reversely degraded if $\eta\leq \frac{1}{2} $.
In the degraded case, $\eta\geq \frac{1}{2}$, the degrading channel  is simply a second beam splitter with
transmissivity 
\begin{align}
\eta'=\frac{1-\eta}{\eta} \,.
\label{eq:etaP}
\end{align}
This is illustrated in Figure~\ref{fig:BSpDegraded}. 
Based on \cite{Guha:08z} (see derivation in the proof of Lemma 3.2 therein), the state of the output mode $\he'$ is the same as that of $\he$. Thereby,
$ \channel^{(2)}_{A\rightarrow E}\equiv \Dset_{B\rightarrow E'} \circ \channel^{(1)}_{A\rightarrow B}$, where $\Dset_{B\rightarrow E'}$ is the bosonic channel corresponding to the second beam splitter.
\end{remark}

\vspace{-0.75cm}

\begin{center}
\begin{figure}[tb]
\includegraphics[scale=0.8,trim={-0.5cm 10.5cm 1cm 10cm},clip]
{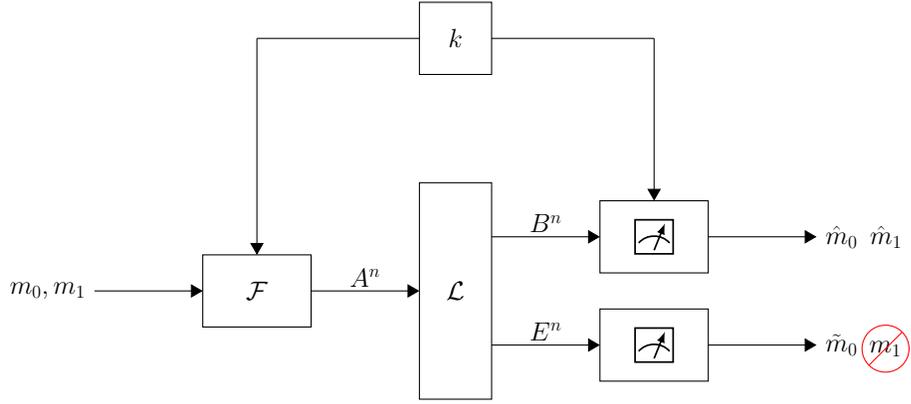} %
\caption{The quantum broadcast channel with confidential messages and key assistance. The sender Alice has the system  $A^n$, and the receivers Bob  and Eve have the systems $B^n$  and $E^n$,  respectively.
Alice and Bob share a random key $k$, which is not known to Eve.
Alice chooses a common message $m_0$ that is intended for both users and a confidential message $m_1$ for Bob.
She encodes the messages by applying the encoding map $\Fset_{  A^n|k}$,
 and transmits $A^n$ over the quantum broadcast channel $\channel_{A\rightarrow B E}$. 
 Bob receives $B^n$ and uses the key $k$ in order to find an estimate of the message pair $(\hm_0,\hm_1)$ by performing a measurement.  
Similarly, Eve measures  $E^n$ and obtains an estimate of the common message $\tm_0$.
 Since Eve is eavesdropping on Bob's confidential message, we require that $m_1$ remains secret from her.
 }
\label{fig:BCkey}
\end{figure}
\end{center}

\subsection{Confidential Coding with and without Key Assistance}
\label{subsec:McodingC}
We define %
a confidential code with and without shared key assistance to transmit classical information over the broadcast channel. 
A common message  is sent to both receivers, Bob and Eve, at a rate $R_0$, and a confidential message  is sent to Bob at a rate $R_1$, while  eavesdropped by Eve. The secret key consists of $nR_K$ random bits, where $R_K$ is a fixed key rate.

\begin{definition} %
\label{def:ClcapacityC}
A $(2^{nR_0},2^{nR_1},n)$ classical  code for the quantum broadcast channel $\channel_{A\rightarrow B E}$ with
confidential messages and key assistance  consists of the following: 
two index sets  $[1:2^{nR_0}]$ and $ [1:2^{nR_1}]$, corresponding to the common message for both users and the confidential message of User 1, respectively, and key index set $[1:2^{nR_K}]$;
 encoding maps $\Fset_{ A^n|k}$ from the product set $[1:2^{nR_0}]\times [1:2^{nR_1}]$ to the input Hilbert space $\Hset_{A^n}$, for $k\in [1:2^{nR_K}]$;	
two collections of decoding POVMs,   $\{\Gamma^{m_0,m_1}_{B^n|k}\}  $, $k\in [1:2^{nR_K}]$, for Bob, and $\{\Xi^{m_0}_{E^n}\}  $ for Eve. %
We denote the code by $(\Fset,\Gamma,\Xi)$.

The communication scheme is depicted in Figure~\ref{fig:BCkey}. 
The sender Alice has the system  $A^n$, and the receivers Bob  and Eve have the systems $B^n$  and $E^n$,  respectively.
A key $k$ is drawn from $[1:2^{nR_K}]$ uniformly at random, and then shared between Alice and Bob. 
Alice chooses a common message $m_0\in [1:2^{nR_0}]$ that is intended for both users and a confidential message $m_1\in [1:2^{nR_1}]$ for Bob, both uniformly at random.
She encodes the messages by applying the encoding map $\Fset_{  A^n|k}$   which results in an input state
$%
\rho^{m_0,m_1,k}_{A^n}= \Fset_{ A^n|k}( m_0,m_1  ) %
$, %
 and transmits the system $A^n$ over $n$ channel uses of $\channel_{A\rightarrow B E}$. Hence, the output state is
\begin{align}
\rho^{m_0,m_1,k}_{ B^n E^n}=\channel^{\otimes n} (\rho^{m_0,m_1,k}_{A^n}) \,.
\end{align}
Eve receives the channel output system $E^n$, and performs a measurement with the POVM 
$\{\Xi^{m_0}_{ E^n}\}  $. From the measurement outcome, she obtains
 an estimate of the common message $\tm_0\in [1:2^{nR_0}]$.
 Similarly, Bob uses the key and performs a POVM $\{\Gamma^{m_0,m_1}_{B^n|k}\}  $ on the output system $B^n$ in order to find an estimate of the message pair $(\hm_0,\hm_1)\in [1:2^{nR_0}]\times [1:2^{nR_1}]$.

The performance of the code is measured in terms of the probability of decoding error and the amount of confidential information that is leaked to Eve.
The conditional probability of error of the code,   
given that the message pair $(m_0,m_1)$ was sent, is given by 
\begin{align}
P_{e|m_0,m_1}^{(n)}(\Fset,\Gamma,\Xi)&= 
1-\frac{1}{2^{nR_K}}\sum_{k=1}^{2^{nR_K}}\trace[  (\Gamma^{m_0,m_1}_{B^n|k}\otimes  \Xi^{m_0}_{E^n})  \rho^{m_0,m_1,k}_{ B^n E^n} ] \,.
\end{align}
The confidential message $m_1$ needs to remain secret from Eve. Thereby, 
the  leakage rate of the code $(\Fset,\Gamma,\Xi)$ is defined as
\begin{align}
s^{(n)}(\Fset)\triangleq   I(M_1;E^n|M_0)_\rho
\,,
\label{eq:elln}
\end{align}
where $M_j$ is a classical random variable that is uniformly distributed over the message index set, $[1:2^{nR_j}]$, for $j=0,1$.

A $(2^{nR_0},2^{nR_1},n,\eps,\delta)$ confidential code satisfies 
$%
\frac{1}{2^{n(R_0+R_1)}}\sum_{m_0,m_1} P_{e|m_0,m_1}^{(n)}(\Fset,\Gamma,\Xi)\leq\eps $ %
and $s^{(n)}(\Fset)\leq \delta$.  %
A rate pair $(R_0,R_1)$, where $R_j\geq 0$, $j=0,1$, is  achievable with key  rate $R_K$  if for every $\eps,\delta>0$ and sufficiently large $n$, there exists a 
$(2^{nR_0},2^{nR_1},n,\eps,\delta)$  code with key assistance. 
 The operational capacity region $\opC_{\text{k-a}}(\channel)$ of the quantum broadcast channel with confidential messages and key assistance
is defined as the set of achievable pairs $(R_0,R_1)$ with a key rate $R_K$. We sometimes refer to $\opC_{\text{k-a}}(\channel)$ as the confidential key-assisted capacity region.
\end{definition}

The confidential capacity region $\opC(\channel)$ without key assistance is defined in a similar manner, as 
as the set of achievable pairs $(R_0,R_1)$ with zero key rate, \ie  $R_K=0$. 

In the bosonic case, it is assumed that the encoder uses a coherent state protocol with an input constraint. That is, the input state is a coherent state $\ket{f_k(m_0,m_1)}$, where the encoding function, $f_k: [1:2^{nR_0}]\times [1:2^{nR_1}]\to \mathbb{C}^n$,
satisfies
$\frac{1}{n}\sum_{i=1}^n |f_{k,i}(m_0,m_1)|^2\leq N_A$, for
$k\in [1:2^{nR_K}]$.

\begin{remark}
We use the standard notation where $R_j$ denotes the private information rate for User $j$, for $j\geq 1$, and $R_0$ denotes the rate of the common messages which is decoded by all receivers. Here, we focus on the case of two receivers, \ie $j\in\{1,2\}$. 
Bob is the name of the receiver of both the common and confidential messages, $m_0$ and $m_1$, while Eve is the second receiver who decodes the common message $m_0$, but also eavesdrops on the confidential message $m_1$. The secrecy requirement is to prevent Eve from decoding  $m_1$.
\end{remark}

\begin{remark}
\label{rem:wiretap}
Taking $R_0=0$, the model reduces to the quantum wiretap channel, where Eve is not required to decode a common message, and she is viewed as a malicious party that is not part of the network. 
In other words, the broadcast channel with confidential messages is a generalization of the wiretap channel.
Alternatively, if one removes the requirement that Eve needs to decode the message $m_0$, then the setting reduces to  an extended wiretap model, in which Alice's message to Bob consists of a public  component and a secret component, as considered in \cite{WildeHsieh:12p,HsiehWilde:09p}. 
The condition in (\ref{eq:elln}) is referred to as strong secrecy (see \eg \cite{SBGPS:21p}). Yet, this requirement is weaker than semantic security or indistinguishability. The results can be extended to stronger security criteria using the methods in \cite{RenesRenner:11p}. %
\end{remark}

\begin{remark}
One may also consider the transmission of quantum information, where the receivers need to recover the state of a pair of quantum
 ``message systems," $\bar{M}_0$ and $\bar{M}_1$. 
However, based on the no-cloning theorem, if Bob can recover the state of the confidential message system $\bar{M}_1$, this automatically guarantees that Eve will not be able to produce this state. Therefore, the quantum capacity region of the broadcast channel with confidential messages is the same as the quantum capacity region without any security requirements.
\label{rem:Qcapacity}
\end{remark}

In the sequel, we also consider a setting with multiple layers of security. To make this introduction brief, the definitions and notations for the layered secrecy  are deferred to Section~\ref{sec:LS}.

\begin{center}
\begin{figure}[tb]
\includegraphics[scale=0.8,trim={-0.5cm 10cm 1cm 10cm},clip]
{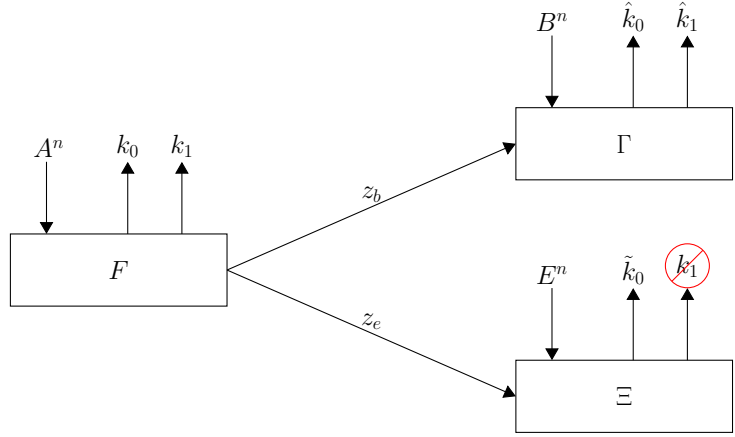} %
\caption{Public and secret key agreement between three terminals.
The terminals Alice, Bob,  and Eve have access to the systems $A^n$, $B^n$,  and $E^n$,  respectively. They distill a pair of public and secret keys from the joint state $\omega_{ABE}^{\otimes n}$, as follows.
Alice produces a pair of public and secret keys, $k_0$ and $k_1$, respectively, by measuring her system using the POVM $\{ F^{k_0,k_1,z_b,z_e}_{A^n} \}$. 
She sends the measurement outcome $z_b$ to Bob, and the measurement outcome $z_e$ to Eve through a public noiseless channel. 
Eve receives $z_e$, and obtains
 a public key $\tk_0$ by performing the measurement  
$\{\Xi^{k_0}_{ E^n|z_e}\}  $ on her system.  The key $k_1$ has to be secret from Eve. 
 Similarly, Bob uses $z_b$ and performs a POVM $\{\Gamma^{k_0,k_1}_{B^n|z_b}\}  $ on his system in order to find a key pair $(\hk_0,\hk_1)$.  }
\label{fig:conferenceKey}
\end{figure}
\end{center}

\subsection{Conference Key Agreement}
\label{subsec:Mcodingk}
Suppose that Alice, Bob, and Eve share a product state $\omega_{ABE}^{\otimes n}$. 
We define %
a  code for the distillation of two keys using their access to this state, with unlimited local operations and classical communication (LOCC). 
A public key  is to be shared with both receivers, Bob and Eve, at a rate $R_0$, and a secret key  is sent to Bob at a rate $R_1$, while  eavesdropped by Eve.

\begin{definition} %
\label{def:ClcapacityEd}
A $(2^{nR_0},2^{nR_1},n)$ key-agreement  code for the distillation of public and secret keys  consists of the following: 
two index sets  $[1:2^{nR_0}]$ and $ [1:2^{nR_1}]$, corresponding to the public key for both users and the secret key for Bob, respectively;
 encoding POVM $\{ F^{k_0,k_1,\nu_{b},\nu_{e}}_{A^n} \}$;	
two collections of decoding POVMs,   $\{\Gamma^{k_0,k_1}_{B^n|z_b}\}  $ for Bob and $\{\Xi^{k_0}_{E^n|z_e}\}  $ for Eve. %
We denote the code by $(F,\Gamma,\Xi)$.

The key-agreement protocol is depicted in Figure~\ref{fig:conferenceKey}. 
The terminals Alice, Bob,  and Eve have access to the systems $A^n$, $B^n$,  and $E^n$,  respectively.
Alice measures her system using the POVM $\{ F^{k_0,k_1,z_b,z_e}_{A^n} \}$. The resulting state is
\begin{align}
\rho_{K_0 K_1 Z_b Z_e B^n E^n}=
\sum_{k_0=1}^{2^{nR_0}} \sum_{k_1=1}^{2^{nR_1}}
\sum_{z_b}\sum_{z_e} \ketbra{k_0,k_1,z_b,z_e} \cdot
\trace_{A^n}\left(  (F^{k_0,k_1,z_b,z_e}_{A^n}\otimes\identity\otimes\identity)\omega^{\otimes n}_{ABE}  \right)  \,.
\end{align}
She sends the measurement outcome $z_b$ to Bob, and the measurement outcome $z_e$ to Eve through a public channel. 
Eve receives $z_e$, and performs a measurement with the POVM 
$\{\Xi^{k_0}_{ E^n|z_e}\}  $. From the measurement outcome, she obtains
 an estimate of the public key $\tk_0\in [1:2^{nR_0}]$.
 Similarly, Bob uses $z_b$ and performs a POVM $\{\Gamma^{k_0,k_1}_{B^n|z_b}\}  $ on the output system $B^n$,  in order to find an estimate of the key pair $(\hk_0,\hk_1)\in [1:2^{nR_0}]\times [1:2^{nR_1}]$.

The performance of the code is based on the probability of distillation error and the amount of secret key that is leaked to Eve.
The average probability of error of the code    is given by 
\begin{align}
P_{e}^{(n)}(F,\Gamma,\Xi)&= 
\Pr\left( \tK_0\neq K_0 \,\,\text{ or }\; 
(\hK_0,\hK_1)\neq (K_0,K_1) \right) \,.
\end{align}
The keys should not be retrieved from the public channel communication.
Furthermore,  $K_1$ needs to remain secret from Eve as well. Thereby, 
we define the leakage rates,
\begin{align}
s_0^{(n)}(F)&\triangleq   I(Z_b,Z_e;K_0) \,,
\\
s_1^{(n)}(F)&\triangleq   I(Z_b,Z_e,E^n;K_1)_\rho \,.
\label{eq:ellnk}
\end{align}

A $(2^{nR_0},2^{nR_1},n,\alpha,\eps,\delta)$  code satisfies 
\begin{align}
\frac{1}{n} H(K_j)&\geq R_j-\alpha \,,
 \\
 P_{e}^{(n)}(F,\Gamma,\Xi)&\leq\eps \,, \\
s^{(n)}_{j}(F)&\leq \delta \,,
\end{align}
for $j=0,1$.
A key-rate pair $(R_0,R_1)$, where $R_j\geq 0$, $j=0,1$, is  achievable if for every $\alpha,\eps,\delta>0$ and sufficiently large $n$, there exists a 
$(2^{nR_0},2^{nR_1},n,\alpha,\eps,\delta)$ key-agreement code. 
 The operational key-agreement capacity region $\Kset(\omega_{ABC})$ for the distillation of  public and  secret keys
is defined as the set of achievable key-rate pairs $(R_0,R_1)$. 
\end{definition}

\begin{remark}
\label{rem:singleK}
If one removes the public key, 
taking $R_0=0$, the model reduces to the
single-user key-agreement setting, as considered by Devetak and Winter \cite{DevetakWinter:05p}. In this setting,  
 Eve is not required to obtain a public key, and she is viewed as a malicious party that tries to get a hold of Bob's secret key. On the other hand, taking $R_1=0$, we distill a public key without eavesdropping, which can be viewed as randomness concentration \cite{AhlswedeCsiszar:93p1,EzzineLabidiBochDeppe:20c}. Nonetheless, we require the public communication to be independent of the distilled key.
\end{remark}

\begin{remark}
The code above is also referred to in the literature as a single-round forward protocol, or a one-way protocol, since we allow Alice to send $(z_b,z_e)$ once to Bob and Eve. In general, there are more complicated key-agreement protocols that include multiple iterations of forward and backward transmissions, from Alice to Bob and/or Eve, and vice versa \cite{DengLong:02p,WMUK:07p,SBCDLP:09p,BlochGunluYener:21p}. 
\end{remark}

\subsection{Related Work}
\label{subsec:Previous}
We briefly review known results for the general broadcast channel with confidential messages in finite dimensions.
Define %
\begin{align}
\mathcal{R}(\channel)=
\bigcup_{ p_{T,X} \,,\; \varphi_{A}^{t,x} } \left\{
\begin{array}{rl}
(R_0,R_1)\,:\; R_0\leq& \min\left( I(T;B)_\rho \,,\; I(T;E)_\rho \right)\\
							 R_1\leq& I(X;B|T)_\rho-I(X;E|T)_\rho
\end{array} \right\} \,,
\label{eq:inCsF}
\end{align}
where the union is over the set of distributions $p_{T,X}$ of two classical auxiliary random variables and collections of
quantum states $\{\varphi_{A}^{t,x}\}$, with 
\begin{align}
\rho_{TX B E}\equiv   \sum_{t,x} p_{T,X}(t,x) \ketbra{t} \otimes \ketbra{x} \otimes \channel(\varphi_{A}^{t,x}) \,.
\end{align}
Notice that here $T$ and $X$ are auxiliary classical variables, which are analogous to $U$ and $V$, respectively, in (\ref{eq:Clbc}). 

The following result on the broadcast channel with confidential messages was recently established by 
 Salek, Hsieh, and Fonollosa \cite{SalekHsieFonollosa:19c,SalekHsieFonollosa:19a}.
\begin{theorem} [see {\cite{SalekHsieFonollosa:19c} \cite[Theorem 3]{SalekHsieFonollosa:19a}}]
\label{theo:CNoSI}
The capacity region of a quantum broadcast channel $\channel_{A\rightarrow BE}$ with confidential messages in finite dimensions is given by 
\begin{align}
\opC(\channel)= \bigcup_{n=1}^\infty \frac{1}{n} \mathcal{R}(\channel^{\otimes n}) \,.
\label{eq:CqNosi}
\end{align}
\end{theorem}
A multi-letter characterization as in (\ref{eq:CqNosi}) is often referred to as a regularized formula.

\begin{remark}
\label{rem:wiretap1}
As pointed out above, the broadcast channel with confidential messages is a generalization of the wiretap channel (see Remark~\ref{rem:wiretap}).
By taking the auxiliary variable $T$ to be null, one obtains the  secrecy rate for the wiretap channel, %
\begin{align}
R^{\text{wiretap}}=
\max_{ p_{X} \,,\; \varphi_{A}^{x} } [I(X;B)_\rho-I(X;E)_\rho] \,.
\end{align}
The secrecy capacity is given by the regularization of the formula above.
We note that when the  channel $\channel_{A\rightarrow BE}$ is reversely degraded (see Subsection~\ref{subsec:Qchannel}), the secrecy capacity is zero, due to the quantum data processing inequality \cite[Theorem 11.5]{NielsenChuang:02b}.
Similarly, if the broadcast channel with confidential messages  is reversely degraded, then confidential information cannot be reliably communicated, \ie $R_1=0$.
\end{remark}

\begin{remark}
\label{rem:privateC}
As pointed out in Remark~\ref{rem:bosonicDeg}, %
if $\channel_{A\rightarrow BE}$ is an isometric broadcast channel, then Eve's system $E$ can be interpreted as Bob's environment.
Then, the point-to-point marginal channel $\channel^{(1)}_{A\rightarrow B}$ is viewed as the main channel, while $\channel^{(2)}_{A\rightarrow E}$ is its complementary. 
In this case, the secrecy capacity of the wiretap channel $\channel_{A\rightarrow BE}$ is also referred to as the \emph{private capacity} of the main channel \cite{Devetak:05p}. This, in turn, is closely related to the \emph{quantum capacity} of this point-to-point channel  (see Remark~\ref{rem:Qcapacity}). In particular, the quantum capacity of a degradable channel $\channel^{(1)}_{A\rightarrow B}$ in finite dimensions  has a single-letter formula and it is identical to the private capacity \cite[Section~13.6.1]{Wilde:17b}.
\end{remark}

\section{Main Results --- Confidential Communication with a Secret Key}
Consider communication of a common message $m_0$ and a confidential message $m_1$ over a broadcast channel $\channel_{A\rightarrow BE}$ with key assistance, as described in Subsection~\ref{subsec:McodingC} and illustrated in Figure~\ref{fig:BCkey}. 
As pointed out in Remark~\ref{rem:wiretap1}, if the broadcast channel $\channel_{A\rightarrow BE}$ is reversely degraded, \ie Bob has a noisier channel than Eve, then secure communication requires that the private rate is zero. However, if Alice and Bob are provided with a secret key, then a positive private rate can be achieved.

We begin with the finite-dimensional case, for which the results are analogous to the classical capacity characterization. 
Then, in the next subsections, we will use those results in order to address the lossy and the pure-loss bosonic broadcast channels.
In the analysis, the main challenge is in the single-letter converse proof for the pure-loss bosonic broadcast channel.

\subsection{Finite Dimensions}
Consider a quantum broadcast channel $\channel_{A\rightarrow BE}$ in finite dimensions.
We give a regularized characterization for the capacity region of the quantum broadcast channel with confidential messages and key assistance. 
Define the rate region
\begin{align}
\mathcal{R}_{\text{k-a}}(\channel,R_K)=
\bigcup_{ p_{T,X} \,,\; \varphi_{A}^{t,x} } \left\{
\begin{array}{lrl}
(R_0,R_1)\,:\; &R_0\leq&  \min\left( I(T;B)_\rho \,,\; I(T;E)_\rho \right) \\
			   &R_1\leq& \min\left( [I(X;B|T)_\rho-I(X;E|T)_\rho]_{+}+R_K \,,\; I(X;B|T)_\rho \right)
\end{array} \right\}\,,
\label{eq:inCskG}
\end{align}
where $[x]_{+}=\max(x,0)$, and the union is over the set of distributions $p_{T,X}$ of two classical auxiliary random variables and collections of
quantum states $\{\varphi_{A}^{t,x}\}$, with 
\begin{align}
\rho_{TX B E}\equiv   \sum_{t,x} p_{T,X}(t,x) \ketbra{t} \otimes \ketbra{x} \otimes \channel(\varphi_{A}^{t,x}) \,.
\end{align}
Before we state the key-assisted capacity theorem, we establish a lemma that allows to compute the region above for a finite-dimensional channel.
 \begin{lemma}
 \label{lemm:PureS}
The union in the the RHS of (\ref{eq:inCskG}) can be exhausted with auxiliary variable cardinalities $|\Tset|\leq |\Hset_A|^4+3$ and $|\Xset|\leq (|\Hset_A|^4+3)(|\Hset_A|^4+2)$.
Furthermore, if $\channel_{A\rightarrow BE}$ is isometric and degraded, then the union can be exhausted by pure states
 $\varphi_A^{t,x}=\ketbra{\phi_A^{t,x}}$.
 \end{lemma}
The proof of Lemma~\ref{lemm:PureS} is given in Appendix~\ref{app:PureS}.
The characterization of the key-assisted capacity region is given in the theorem below.
\begin{theorem} 
\label{theo:KAqbc}
The capacity region of a quantum broadcast channel $\channel_{A\rightarrow BE}$ with confidential messages and key assistance in finite dimensions is given by 
\begin{align}
\opC_{\text{k-a}}(\channel)= \bigcup_{n=1}^\infty \frac{1}{n} \mathcal{R}_{\text{k-a}}(\channel^{\otimes n},nR_K) \,.
\label{eq:CqNosik}
\end{align}
\end{theorem}
Observe that by taking a zero key rate, \ie $R_K=0$, we recover the unassisted capacity region in 
Theorem~\ref{theo:CNoSI},
due to Salek \etal \cite{SalekHsieFonollosa:19a,SalekHsieFonollosa:19c}.
The proof of Theorem~\ref{theo:KAqbc} is given in Appendix~\ref{app:KAqbc}.

\begin{remark}
In the achievability proof in Appendix~\ref{app:KAqbc},
we use a similar approach as originally used by Yamamoto \cite{Yamamoto:97p} and Kange and Liu \cite{KangLiu:10c}.
We apply the one-time pad coding scheme. We use rate-splitting in order to combine between the one-time pad coding scheme and the unassisted confidential coding scheme due to Salek \etal \cite{SalekHsieFonollosa:19a,SalekHsieFonollosa:19c}.
That is, the private message rate is decomposed as  $R_1=R_{1\mathrm{k}}+R_{1\mathrm{c}}$, where the rates $R_{1\mathrm{k}}$ and $R_{1\mathrm{c}}$ correspond to the key-assisted encryption and the unassisted confidential code respectively. As can be seen in the proof, the unassisted rate must satisfy $R_{1\mathrm{c}}\leq [I(X;B|T)_\rho-I(X;E|T)_\rho]_{+}$.
Therefore, if the quantum broadcast channel is reversely degraded, then $R_{1\mathrm{c}}=0$ and the confidentiality relies solely on the one-time pad cypher. The regularized converse proof is analogous to the classical proof in \cite{SchaeferKhistiPoor:18p}.
\end{remark}

\subsection{Lossy Bosonic Channel}
We establish an inner bound on the capacity region of the lossy bosonic broadcast channel with confidential messages. Denote the lossy bosonic broadcast channel by $\channel_{\,\text{lossy}}$.
Note that the input constraint, the channel transmissivity, and the noise mean photon number, \ie
$N_A$, $\eta$, and $N_C$, and are all fixed in this model.
Define $\mathcal{R}_{\text{in}}(\eta,R_K)$ as follow.
If $\eta\geq \frac{1}{2}$, let
\begin{multline}
\mathcal{R}_{\text{in}}(\eta,R_K)
=\\
\bigcup_{ 0\leq \beta\leq 1} \left\{
\begin{array}{rl}
(R_0,R_1)\,:\; R_0\leq&  g((1-\eta) N_A+\eta N_C)-g((1-\eta) \beta N_A+\eta N_C) \\
							 R_1\leq& g(\eta \beta N_A+(1-\eta) N_C)-g((1-\eta) N_C)-[g((1-\eta) \beta N_A+\eta N_C)-g(\eta N_C)]+R_K \\
				R_1\leq& g(\eta \beta N_A+(1-\eta) N_C)-g((1-\eta) N_C)
\end{array} \right\} \,,
\end{multline}
otherwise, if $\eta< \frac{1}{2}$, let
\begin{align}
\mathcal{R}_{\text{in}}(\eta,R_K)
=
\bigcup_{ 0\leq \beta\leq 1} \left\{
\begin{array}{rl}
(R_0,R_1)\,:\; R_0\leq&  g(\eta N_A+(1-\eta) N_C)-g(\eta \beta N_A+(1-\eta) N_C) \\
				R_1\leq& \min\left( g(\eta \beta N_A+(1-\eta) N_C)-g((1-\eta) N_C)\,,\; R_K \right)
\end{array} \right\} \,,
\end{align}
where $g(N)$  is the entropy of the thermal state $\tau(N)$ (see (\ref{eq:tau})), namely,
\begin{align}
    g(N)=
    \begin{cases}
    (N+1)\log(N+1)-N\log(N) & N>0\\
    0 & N=0\,.
    \end{cases}
\end{align}
The subscript `in' stands for `inner bound'.

\begin{theorem}
\label{theo:LossyC}
The capacity region of the lossy bosonic broadcast channel with confidential messages and key assistance satisfies
\begin{align}
    \opC_{\text{k-a}}(\channel_{\,\text{lossy}})
    \supseteq \mathcal{R}_{\text{in}}(\eta,R_K) \,.
\end{align}
\end{theorem}
Before we give the proof, we establish an  achievable region without key assistance as a consequence.
\begin{corollary}
The capacity region of the lossy bosonic broadcast channel with confidential messages without key assistance satisfies
$
    \opC(\channel_{\,\text{lossy}})
    \supseteq \mathcal{R}_{\text{in}}(\channel_{\,\text{lossy}},0)
$. 
Specifically, if $\eta\geq\frac{1}{2}$, then
\begin{multline}
    \opC(\channel_{\,\text{lossy}})
    \supseteq \\
   \bigcup_{ 0\leq \beta\leq 1} \left\{
\begin{array}{rl}
(R_0,R_1)\,:\; R_0\leq&  g((1-\eta) N_A+\eta N_C)-g((1-\eta) \beta N_A+\eta N_C) \\
							 R_1\leq& g(\eta \beta N_A+(1-\eta) N_C)-g((1-\eta) N_C)-[g((1-\eta) \beta N_A+\eta N_C)-g(\eta N_C)] 
\end{array} \right\} \,.
\end{multline}
If $\eta<\frac{1}{2}$, 
\begin{align}
    \opC(\channel_{\,\text{lossy}})
    &\supseteq 
 \left\{
(R_0,0)\,:\; R_0\leq  g(\eta N_A+(1-\eta) N_C)-g((1-\eta) N_C)  
 \right\} \,.
\end{align}
\end{corollary}
Notice that as pointed out in Remark~\ref{rem:wiretap1}, if the broadcast channel is reversely degraded, then we cannot send confidential messages without key assistance, \ie $R_1=0$.

To prove Theorem~\ref{theo:LossyC},
we extend the finite-dimension result in Theorem~\ref{theo:CNoSI} to the bosonic channel with infinite-dimension Hilbert spaces based on the discretization limiting argument by Guha \etal \cite{GuhaShapiroErkmen:07p}. 
Further discussion and justification for this argument are given in Subsection~\ref{subsec:Discretization}.
If one ignores the input constraint, then based on Theorem~\ref{theo:CNoSI}, the region $\mathcal{R}(\channel_{\text{lossy}})$ is achievable, \ie
$\opC(\channel_{\text{lossy}})\supseteq \mathcal{R}(\channel_{\text{lossy}})$.
Suppose that $\eta\geq \frac{1}{2}$.
Since the bosonic channel is degraded, 
$I(T;B)\geq I(T;E)_\rho$ for every input state by the quantum data processing inequality. Thus, we obtain the following inner bound,
\begin{align}
    \opC(\channel_{\text{lossy}})\supseteq   \left\{
 \begin{array}{rl}
 (R_0,R_1)\,:\; R_0\leq& I(T;E)_\rho\\
 							 R_1\leq&  I(X;B|T)_\rho-I(X;E|T)_\rho+R_K \\
 							 R_1\leq& I(X;B|T)_\rho 
 \end{array} \right\} \,,
\end{align}
where the auxiliary variables $T$ and $X$ can be chosen arbitrarily. Given the input constraint, we need to add the restriction  
$\mathbb{E}(|X|^2)\leq N_A$.

Then, set the input to be a coherent state with $X=T+Q$, where
 $T$ and $Q$ are  independent complex Gaussian random variables, 
\begin{align}
T &\sim \mathcal{N}_{\mathbb{C}}(0,(1-\beta)\frac{N_A}{2})\\
Q &\sim \mathcal{N}_{\mathbb{C}}(0,\beta\frac{N_A}{2})
\end{align}
for some $\beta\in [0,1]$.
Then, $X$ is distributed according to
$%
\sim \mathcal{N}_{\mathbb{C}}(0,\frac{N_A}{2}) %
$. %
The distribution of $X$ given $T=t$ is $\mathcal{N}_{\mathbb{C}}(t,\beta\frac{N_A}{2})$, hence
\begin{align}
I(T;E)_\rho&=H(E)_\rho-H(E|T)_\rho=g((1-\eta) N_A+\eta N_C)-g((1-\eta) \beta N_A+\eta N_C) \\
I(X;B|T)_\rho&= H(B|T)_\rho - H(B|X)_\rho=g(\eta \beta N_A+(1-\eta) N_C)-g((1-\eta) N_C)\\
I(X;E|T)_\rho&= H(E|T)_\rho - H(E|X)_\rho=g((1-\eta) \beta N_A+\eta N_C)-g(\eta N_C)
\end{align}
Note that $I(X;B|T)_\rho\geq I(X;E|T)_\rho$ if and only if $\eta\geq\frac{1}{2}$. The proof for the reversely degraded case, \ie $\eta<\frac{1}{2}$, follows similar arguments, and is thus omitted.
This completes the achievability proof.

\begin{remark}
As in the classical case, the choice above has the interpretation of ``a superposition coding scheme" \cite{ElGamalKim:11b}. For simplicity, consider the unassisted setting, \ie with $R_K=0$. The scheme consists of a collection of sequences $t^n(m_0)$ and 
$q^n(m_0,m_1)$, for $m_0\in [1:2^{nR_0}]$ and $m_1\in [1:2^{nR_1}]$. 
The sequences $t^n(m_0)$ are called cloud centers, while $x^n(m_0,m_1)=t^n(m_0)+q^n(m_0,m_1)$ are thought of as satellites.
Hence, the common message $m_0$ is an index of the cloud center, and the confidential message $m_1$ indicates the cloud satellite.
Imagine that each cloud center is a point located at a distance of $\sqrt{n(1-\beta) N_A}-\eps$ from the origin.
Furthermore, from each cloud center $t^n(m_0)$, emerges a cloud vector $q^n(m_0,m_1)$ of length $\sqrt{n\beta N_A}$.
Then, the satellites of each cloud have an $\ell_2$-norm of at most 
$\sqrt{nN_A}-\eps$, by the triangle inequality. In order to ensure security, the radius $q^n(m_0,m_1)$ is chosen at random from a bin that consists of 
$2^{n(I(X;E|T)_\rho+\delta)}$ sequences. Therefore, if the rate pair $(R_0,R_1)$ is in $\mathcal{R}_{\text{in}}(\channel_{\text{lossy}},0)$, then Eve can recover the cloud center chosen by Alice, but she cannot determine which satellite was used. Whereas, Bob can decode both the center and the satellite.
\end{remark}

\begin{center}
\begin{figure}[tb]
\includegraphics[scale=0.6,trim={-5cm 0 1cm 0},clip
]
{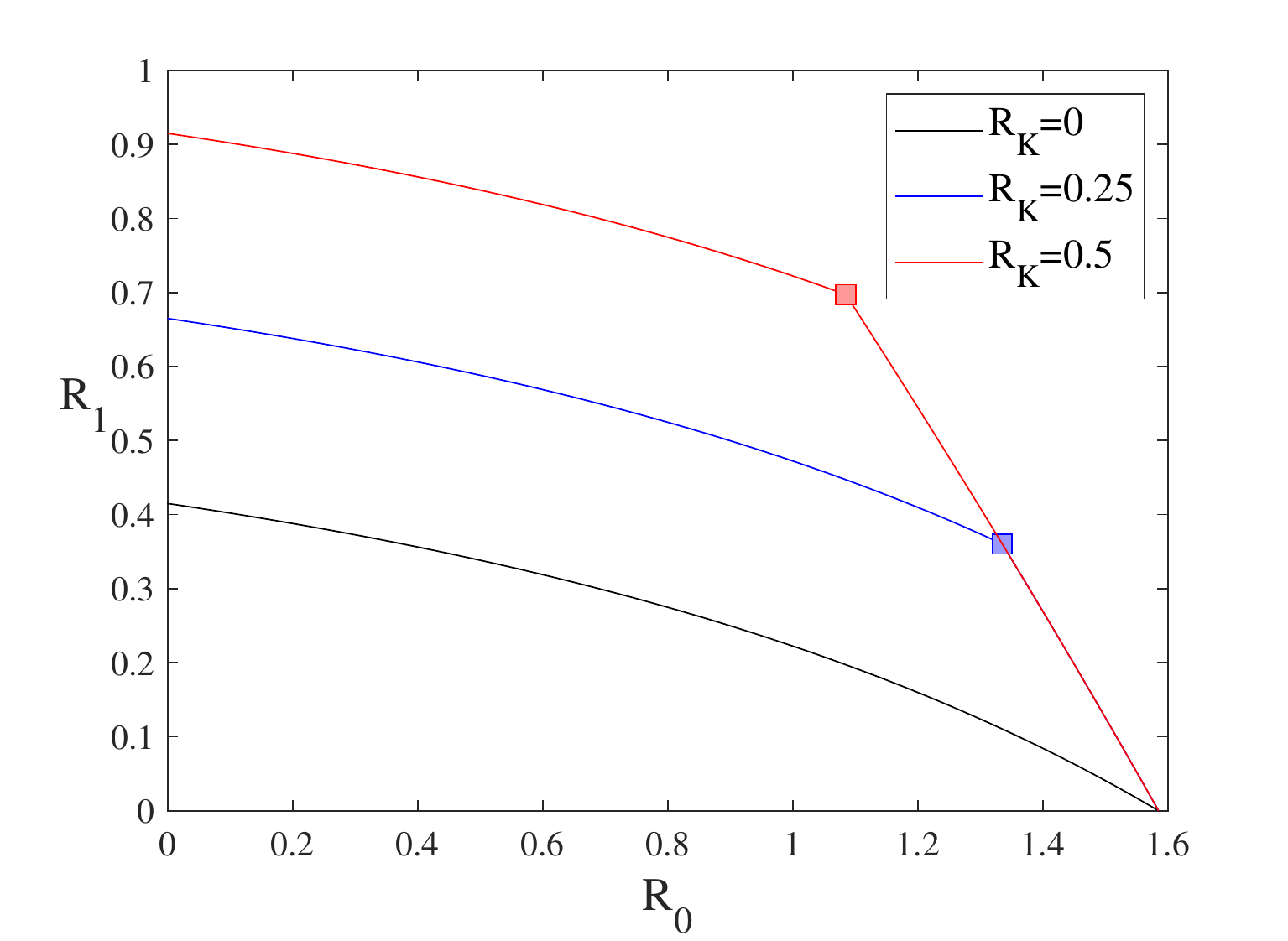} %
\caption{The capacity region of the pure-loss bosonic broadcast channel with confidential messages and key assistance, given the transmissivity $\eta=0.6$, and  input constraint $N_A=5$. The black, blue, and red lines correspond to the key rates $R_K=0$, $R_K=0.25$, and $R_K=0.5$, respectively.
The squares mark a transition (``breaking point") in each region.
For  $R_K=0.5$ (the red curve), the breaking point is $(R_0,R_1)=(1.085,0.697)$, which corresponds to $\beta=0.207$. For low common rates, %
$R_0<1.085$, 
the shared key is fully used to enhance the communication rates, whereas for higher rates $R_0>1.085$, the key is only partially used due to the limitation of Bob's channel to decode the messages.
}
\label{fig:BosonicKey}
\end{figure}
\end{center}

\subsection{Pure-Loss Bosonic Channel}
\label{subsec:PureLmainC}
For the pure-loss bosonic broadcast channel, in which the noise mode is in the vacuum state, we 
determine the capacity region exactly, under the assumption of the  minimum output-entropy conjecture. This long-standing conjecture is known to hold in special cases \cite{Palma:19p}. As mentioned above, the main technical challenge is in the single-letter converse proof, which requires the conjecture.
Denote the channel by $\channel_{\text{pure-loss}}$.
\begin{conjecture}[see \cite{GuhaShapiro:07c}]
\label{con:MinOutStrong}
Let the noise modes $\{ \hc_i   \}_{i=1}^n$ be in a product state $\rho_{C^n}=\ketbra{0}^{\otimes n}$ of $n$ vacuum states, and assume that 
$H(A^n)_\rho=n g( N_A)$. Then,
\begin{align}
 H(B^n)_\rho \geq n g (\eta N_A) \,.
\end{align}
\end{conjecture}

\begin{theorem}
\label{theo:PlossCf}
Assume that Conjecture~\ref{con:MinOutStrong} holds.
Then, the capacity region of the pure-loss bosonic broadcast channel with confidential messages is as follows. If $\eta\geq \frac{1}{2}$, then 
\begin{align}
    \opC(\channel_{\text{pure-loss}})=\bigcup_{ 0\leq \beta\leq 1} \left\{
\begin{array}{rl}
(R_0,R_1)\,:\; R_0\leq&  g((1-\eta) N_A)-g((1-\eta) \beta N_A) \\
							 R_1\leq& g(\eta \beta N_A)-g((1-\eta) \beta N_A)+R_K
							 \\
							 R_1\leq& g(\eta \beta N_A)
\end{array} \right\} \,.
\end{align}
Otherwise, if $\eta<\frac{1}{2}$,
\begin{align}
    \opC(\channel_{\text{pure-loss}})=\bigcup_{ 0\leq \beta\leq 1} \left\{
\begin{array}{rl}
(R_0,R_1)\,:\; R_0\leq&  g((1-\eta) N_A)-g((1-\eta) \beta N_A) \\
							 R_1\leq& \min\left( g(\eta \beta N_A)\,,\; R_K\right)
\end{array} \right\} \,.
\end{align}
\end{theorem}
The capacity region of the pure-loss bosonic broadcast channel is depicted in Figure~\ref{fig:BosonicKey} for different key values, transmissivity $\eta=0.6$, and  input constraint $N_A=5$. 
 The black, blue, and red lines correspond to the key rates $R_K=0$, $R_K=0.25$, and $R_K=0.5$, respectively.
The squares mark the phase transition (``breaking point") in each region.
For  $R_K=0.5$ (the red curve), the breaking point is $(R_0,R_1)=(1.085,0.697)$, which corresponds to $\beta=0.207$. For low common rates, %
$R_0<1.085$, 
the shared key is fully used to enhance the communication rates, whereas for higher rates $R_0>1.085$, the key is only partially used due to the limitation of Bob's channel to decode the messages.
In general, the breaking point corresponds to the value $\beta_0$ such that 
\begin{align}
    g((1-
\eta)\beta_0 N_A)=R_K \,.
\end{align}

Consider the first part of Theorem~\ref{theo:PlossCf}. The direct part follows immediately from Theorem~\ref{theo:LossyC}, by taking $N_C\rightarrow 0$.
To show the converse part,
we combine the arguments of Guha and Shapiro \cite{GuhaShapiro:07c} for the pure-loss bosonic channel  with the methods of Smith \cite{Smith:08p} for the degraded wiretap channel.
The proof requires the strong minimum output-entropy conjecture. That is, we assume that Conjecture~\ref{con:MinOutStrong} holds. 

Moving the converse proof, let Alice and Bob share a random key $k$ uniformly distributed over $[1:2^{nR_K}]$. Suppose that Alice chooses $m_0$ and $m_1$ uniformly at random, and  prepares an input state $\rho^{m_0,m_1,k}_{ A^n}$. 
After Alice sends the system $A^n$ through the channel,  the output state is
$\rho_{B^n E^n}=\frac{1}{2^{n(R_0+R_1+R_K)}}\sum_{m_0=1}^{2^{nR_0}} \sum_{m_1=1}^{2^{nR_1}}
\sum_{k=1}^{2^{nR_K}}
\channel_{A^n\rightarrow B^n E^n}(\rho^{m_0,m_1,k}_{ A^n}) $.
Then, Bob and Eve perform  decoding POVMs $\Gamma^{m_0,m_1}_{B^n|k}$ and $\Xi^{m_0}_{E^n}$, respectively.
Consider a sequence of codes $(\Fset_n,\Lambda_n,\Gamma_n)$ such that the average probability of error and the leakage tend to zero, hence
the error probabilities $\prob{ \hM_0\neq M_0 }$, $\prob{ (\hM_0,\hM_1)\neq (M_0,M_1)}$, $\prob{ \hM_1\neq M_1 |M_0}$,   are bounded by some
$\alpha_n$ which tends to zero as $n\rightarrow \infty$.
By Fano's inequality \cite{CoverThomas:06b}, it follows that%
\begin{align}
H(M_0|\tM_0) \leq n\eps_n
\label{eq:Hm0u}
\\
H(M_1|\hM_1,M_0) \leq n\eps_n'
\label{eq:Hm1u}
\end{align}
where $\eps_n,\eps_n'$ tend to zero as $n\rightarrow\infty$.
Since the leakage tends to zero, we also have
\begin{align}
I(M_1;E^n|M_0)_\rho\leq n\delta_n
\label{eq:Lconverse}
\end{align}
where $\delta_n$ tends to zero as $n\rightarrow\infty$.

First, we show the following multi-letter upper bounds,
\begin{align}
R_0&\leq \frac{1}{n} I(M_0;E^n )_{\rho}+\eps_n
\label{eq:multiUp1}
\\
R_1&\leq \frac{1}{n}  H(B^n|M_0,K)_\rho+\eps_n' \,.
\label{eq:multiUp2}
\\
R_1&\leq \frac{1}{n}  [H(B^n|M_0,K)_\rho-H(E^n|M_0,K)_\rho]+R_K+\delta_n+\eps_n' \,.
\label{eq:multiUp3}
\end{align}
Indeed, the common rate is bounded as 
\begin{align}
nR_0&= H(M_0)=I(M_0;\tM_0)_{\rho}+H(M_0|\tM_0) 
\nonumber\\
&\leq I(M_0;\tM_0)_{\rho}+n\eps_n \nonumber\\
&\leq I(M_0;E^n )_{\rho}+n\eps_n %
\label{eq:multiUp1a}
\end{align}
where the first inequality follows from  (\ref{eq:Hm0u}), and the last inequality %
follows from  the Holevo bound due to data processing inequality (see  \cite[Theorem 12.1]{NielsenChuang:02b}). %

Similarly, the private rate is bounded as
\begin{align}
   nR_1 &\leq I(M_1;B^n,K|M_0)_\rho+n\eps_n' 
   \nonumber\\
   &=I(M_1;B^n|M_0,K)_\rho+n\eps_n'
   \label{eq:nR1upp3a}
   \\
   &\leq H(B^n|M_0,K)_\rho+n\eps_n'
   \label{eq:nR1upp3}
\end{align}
where the equality follows from  the  statistical independence between the key and the messages, and the last inequality holds since $M_0$, $M_1$, $K$ are classical.
Then, (\ref{eq:nR1upp3a}) implies that
\begin{align} 
nR_1 &\leq I(M_1;B^n|M_0,K)_\rho-I(M_1;E^n|M_0)_\rho+n\delta_n+n\eps_n'
\label{eq:nR1upp0}
\end{align}
since we have seen in (\ref{eq:Lconverse}) that $-I(M_1;E^n|M_0)_\rho+ n\delta_n\geq 0$ because of the leakage requirement.
Using the chain rule, we have
\begin{align}
    I(M_1;E^n|M_0)_\rho&=
    I(M_1;E^n,K|M_0)_\rho-I(M_1;K|E^n,M_0)_\rho
    \nonumber\\
    &\geq I(M_1;E^n,K|M_0)_\rho-H(K)
    \nonumber\\
    &= I(M_1;E^n|M_0,K)_\rho-nR_K
    \label{eq:nR1upp0b}
\end{align}
where the last two lines follow because the key is classical and uniformly distributed over $[1:2^{nR_K}]$. Inserting (\ref{eq:nR1upp0b}) into the bound on the private rate in (\ref{eq:nR1upp0}), we obtain
\begin{align}
    nR_1 &\leq I(M_1;B^n|M_0,K)_\rho-I(M_1;E^n|M_0,K)_\rho+nR_K+n(\delta_n+\eps_n')
    \label{eq:nR1upp0a}
\end{align}

Now, consider a spectral decomposition of the input state,
\begin{align}
    \rho^{m_0,m_1,k}_{A^n}=
    \int  %
    p_{Y|M_0,M_1,K}(y|m_0,m_1,k) \ketbra{\phi_{A^n}^{y,m_0,m_1,k}}  \, dy 
\end{align}
where $Y$ is an ``index" over the continuous ensemble  $\{\ket{\phi_{A^n}^{y,m_0,m_1,k}}\}$, and $p_{Y|M_0,M_1,K}(y|m_0,m_1,k)$ is a conditional probability density function.
Then, augmenting $Y$, we obtain the extended output state,
\begin{align}
    \rho_{ Y  B^n E^n}^{m_0,m_1,k}
    &=
    \int p_{Y|M_0,M_1,K}(y|m_0,m_1,k)
    \ketbra{y}_Y\otimes (U^{\channel})^{\otimes n}
    \ketbra{\phi_{B^n E^n}^{y,m_0,m_1,k}} (U^{\channel})^{\otimes n}  \, dy
\end{align}
where $U^{\channel}$ is the isometry corresponding to the bosonic broadcast channel. 
By the chain rule,
\begin{align}
    I(M_1;B^n|M_0,K)_\rho-I(M_1;E^n|M_0,K)_\rho &=
    I(M_1,Y;B^n|M_0,K)_\rho-I(M_1,Y;E^n|M_0,K)_\rho 
    \nonumber\\&
    -[ I(Y;B^n|M_0,M_1,K)_\rho-I(Y;E^n|M_0,M_1,K)_\rho]
    \nonumber\\&
    \leq I(M_1,Y;B^n|M_0,K)_\rho-I(M_1,Y;E^n|M_0,K)_\rho 
\end{align}
where the last inequality holds since the bosonic broadcast channel is degraded and $I(Y;B^n|M_0,M_1,K)_\rho\geq I(Y;E^n|M_0,M_1,K)_\rho$ 
by the quantum data processing inequality. 

Notice that given $M_j=m_j$, $K=k$, and $Y=y$, the joint state of $(B^n,E^n)$ is pure. Therefore,
$H(B^n|M_0,M_1,K,Y)_\rho=H(E^n|M_0,M_1,K,Y)_\rho$, and
\begin{align}
   I(M_1,Y;B^n|M_0,K)_\rho-I(M_1,Y;E^n|M_0,K)_\rho = H(B^n|M_0,K)_\rho-H(E^n|M_0,K)_\rho \,.
\end{align}
Therefore, (\ref{eq:nR1upp0a}) becomes
\begin{align}
    nR_1\leq H(B^n|M_0,K)_\rho-H(E^n|M_0,K)_\rho+n(R_K+\delta_n+\eps_n') \,.
    \label{eq:multiUp2a}
\end{align}
We have thus proved  the multi-letter upper bounds (\ref{eq:multiUp1})-(\ref{eq:multiUp3}), which immediately follow from 
(\ref{eq:multiUp1a}), (\ref{eq:nR1upp3}), and (\ref{eq:multiUp2a}), respectively, by dividing both sides of each inequality by $n$.

To prove the single-letter converse part, we proceed as follows.
Since the thermal state maximizes the quantum entropy over all states with the same first and second moments \cite{WolfGiedkeCirac:06p}, 
\begin{align}
H(B^n|M_0,K)_\rho &\leq \sum_{i=1}^n H(B_i|M_0,K)_\rho
\nonumber\\
&\leq \sum_{i=1}^n g( N_{B_i})
\label{eq:HBnBound}
\end{align}
where $N_{B_i}$ is the mean photon number that corresponds to the $i^{\text{th}}$ output. As $g(\cdot)$ is  concave and monotonically increasing,  
\begin{align}
\frac{1}{n}\sum_{i=1}^n g( N_{B_i}) &\leq g\left( \frac{1}{n}\sum_{i=1}^n  N_{B_i} \right)
\nonumber\\
& \leq  g(\eta N_A ) \,.
\end{align}
Together with (\ref{eq:HBnBound}), this implies $\frac{1}{n} H(B^n|M_0,K)_\rho \leq g(\eta N_A )$. Thereby, there exists
$0\leq \beta\leq 1$ such that 
\begin{align}
\frac{1}{n} H(B^n|M_0,K)_\rho = g(\eta \beta N_A ) \,.
\label{eq:EntBnCond1b}
\end{align}
Recall, from Remark~\ref{rem:bosonicDeg}, that  Eve's degraded state can be obtained as the output of a pure-loss bosonic channel, where the input is Bob's state, and the transmissivity for this degrading channel is 
$\eta'=\frac{1-\eta}{\eta}$  (see Figure~\ref{fig:BSpDegraded}). %
Thereby, assuming Conjecture~\ref{con:MinOutStrong} holds, we can deduce from (\ref{eq:EntBnCond1b}) that
\begin{align}
\frac{1}{n} H(E^n|M_0,K)_\rho \geq g(\eta'\cdot \eta \beta N_A )= g((1-\eta)\beta N_A) \,.
\label{eq:EntEnCond1}
\end{align}

By similar considerations,
\begin{align}
\frac{1}{n} H(E^n)_\rho\leq \frac{1}{n} \sum_{i=1}^n g(N_{E_i})\leq g\left( \frac{1}{n}\sum_{i=1}^n  N_{E_i} \right) \leq g( (1-\eta)N_A ) \,.
\end{align}
Thus,
\begin{align}
R_0-\eps_n &\stackrel{\eqref{eq:multiUp1}}{\leq} 
\frac{1}{n} I(M_0,K;E^n)_\rho
\nonumber\\
&=\frac{1}{n} [H(E^n)_\rho-H(E^n|M_0,K)_\rho
\nonumber\\
&\stackrel{(\ast)}{\leq} g( (1-\eta)N_A ) -g( (1-\eta)\beta N_A )
\\
\,
\\
R_1-\eps_n' &\stackrel{\eqref{eq:multiUp2}}{\leq} 
\frac{1}{n} H(B^n|M_0,K)_\rho
\nonumber\\
&\stackrel{(\ast)}{=} g( \eta\beta N_A ) 
\intertext{and}
R_1-\delta_n-\eps_n' &\stackrel{\eqref{eq:multiUp3}}{\leq} 
\frac{1}{n} [H(B^n|M_0,K)_\rho-H(E^n|M_0,K)_\rho]+R_K
\nonumber\\
&\stackrel{(\ast)}{\leq} g( \eta\beta N_A ) -g( (1-\eta)\beta N_A )+R_K  
\end{align}
where $(\ast)$ follow from (\ref{eq:EntBnCond1b})-(\ref{eq:EntEnCond1}).
\qed

\section{Main Results -- Key Agreement}
Consider the distillation of a public key and a secret key between Alice, Bob, and Eve, using a correlated state $\omega_{ABE}^{\otimes n}$, as described in Subsection~\ref{subsec:Mcodingk}, and illustrated in Figure~\ref{fig:conferenceKey}. This source model is rather different compared to the confidential channel model in the previous section. Yet, we will point out the connection between them in Corollary~\ref{coro:distillationF}, and in particular, the relation to the bosonic broadcast channel.

 We characterize the key-agreement capacity region for the case where $A$, $B$, and $E$ have finite dimensions
Define a key-rate region,
\begin{align}
\mathsf{K}(\omega_{ABE})=
\bigcup_{ \Lambda_A \,,\; p_{T_0,T_1|X}  } \left\{
\begin{array}{lrl}
(R_0,R_1)\,:\; &R_0\leq&  \min\left( I(T_0;B)_\omega \,,\; I(T_0;E)_\omega \right) \\
			   &R_1\leq& [I(X;B|T_0,T_1)_\omega-I(X;E|T_0,T_1)_\omega]_{+}
\end{array} \right\}\,,
\label{eq:inCskGd}
\end{align}
where $[x]_{+}=\max(x,0)$, and the union is over the set of POVMs $\Lambda_A=\{\Lambda_A^x\}_{x\in\Xset}$ and conditional
distributions $p_{T_0,T_1|X}$, with 
\begin{align}
\omega_{T_0 T_1 X B E}\equiv   \sum_{t_0,t_1,x} p_{T_0,T_1|X}(t_0,t_1|x) \ketbra{t_0} \otimes \ketbra{t_1} \otimes \ketbra{x} \otimes \trace_A\left( (\Lambda_A^x\otimes\identity\otimes \identity) \omega_{ABE} \right) \,.
\end{align}
Since any measurement can be extended to a projective measurement \cite[Section 2.2.8]{NielsenChuang:02b}, the auxiliary variables alphabets can be restricted to $|\Xset|\leq (|\Hset_A| |\Hset_B| |\Hset_E|)^4$, $|\Tset_0|\leq |\Xset|+3$,  and  $|\Tset_1|\leq |\Tset_0||\Xset|$, by the same arguments as in the proof of Lemma~\ref{lemm:PureS} in Appendix~\ref{app:PureS}. Hence, the formula above is in principle computable.

The key-agreement theorem is given below.
\begin{theorem}
\label{theo:distillationF}
The key-agreement capacity region for the distillation of a public key and a secret key from $\omega_{ABE}$ in finite dimensions is given by
\begin{align}
    \Kset(\omega_{ABC})=\bigcup_{n=1}^\infty\frac{1}{n}\mathsf{K}(\omega_{ABC}^{\otimes n}) \,.
\end{align}
\end{theorem}
The proof of Theorem~\ref{theo:distillationF} is given in Appendix~\ref{app:distillationF}. We note that in the multi-letter capacity result, the auxiliary variable $T_1$ is not necessary, as can be seen in the converse proof. 
However, it %
emerges as a result of the single letterization for special cases, including a classical channel.
Although, in the degraded case, $T_1$ is not necessary even for a single-letter characterization.
 We observe the following relation with the broadcast channel with confidential messages.
\begin{corollary}
\label{coro:distillationF}
Let $\channel_{A\rightarrow BE}$ be a degraded broadcast channel.
Then, 
\begin{align}
     \bigcup_{\omega_{ABE}\,:\; \omega_{BE}=\channel_{A\rightarrow BE}(\omega_A)} \mathcal{K}(\omega_{ABC})=\lim_{n\rightarrow\infty}\frac{1}{n}\mathcal{R}(\channel^{\otimes n}) \,,
\end{align}
where $\mathcal{R}(\cdot) $ is as defined in (\ref{eq:inCsF}).
Thus, The capacity region of a degraded broadcast channel $\channel_{A\rightarrow BE}$ with confidential messages satisfies
\begin{align}
     \opC(\channel)=\bigcup_{\omega_{ABE}\,:\; \omega_{BE}=\channel_{A\rightarrow BE}(\omega_A)} \mathcal{K}(\omega_{ABC}) \,. 
\end{align}
\end{corollary}
Corollary~\ref{coro:distillationF} follows from Theorem~\ref{theo:distillationF} in a straightforward manner.
Indeed, consider the first part of the corollary.
Then, the direct part follows by taking $X=T_1$. As for the converse part, we note that for a degraded channel, $I(T_1;B|T_0)_\omega\geq I(T_1;E|T_0)_\omega$, by the data processing inequality. Thus,
\begin{align}
    I(X;B|T_0,T_1)_\omega-I(X;E|T_0,T_1)_\omega&\leq 
    I(X;B|T_0,T_1)_\omega-I(X;E|T_0,T_1)_\omega+[I(T_1;B|T_0)_\omega- I(T_1;E|T_0)_\omega]
    \nonumber\\
    &=I(\tX;B|T_0)_\omega-I(\tX;E|T_0)_\omega
\end{align}
where we have defined $\tX\equiv (X,T_0) $. The second part of the corollary follows from Theorem~\ref{theo:CNoSI}.
Based on this corollary,  the key-agreement capacity region is included within the corresponding confidential capacity region without key assistance.
In particular, the key-agreement capacity region for thermal states that are associated with a pure-loss bosonic channel is the same as %
the  confidential capacity region in   Theorem~\ref{theo:PlossCf}.

\section{Layered Secrecy}
\label{sec:LS}
As mentioned in the introduction, the quantum broadcast channel with layered decoding and secrecy is a generalization of the degraded broadcast channel with confidential messages \cite{LyLiuBlankenship:12p,ZouLiangLaiShamai:15p1,ZouLiangLaiPoorShamai:15p}. 
The model describes a network in which the users have different credentials to access confidential information. 
Zou \etal \cite{ZouLiangLaiPoorShamai:15p} give the practical example of %
an agency WiFi network, in which a user is allowed to receive files up to a certain security clearance but should be kept ignorant of classified files that require a higher security level. 
As pointed out in \cite{ZouLiangLaiPoorShamai:15p}, the agency can set the channel quality on a clearance basis by assigning more communication resources
to users with a higher clearance. 
As before, we begin with the finite-dimensional case, and then consider the bosonic channel. We will describe a  bosonic network with three receivers, where the channel is formed by a serial connection of two beam splitters, as illustrated in Figure~\ref{fig:BSpLayeredS}.
By extending the finite-dimension results, we will derive an achievable layered-secrecy region for the pure-loss bosonic broadcast channel.

Consider a channel $\channel_{A\rightarrow B E_1 E_2}$ with three receivers, Bob, Eve 1, and Eve 2. 
The sender, Alice, sends three messages, $m_0$, $m_1$, and $m_2$. 
The information has different layers of confidentiality. 
The message $m_0$ is a common message that is intended for all three receivers.
In the next layer, the confidential message $m_1$ is decoded by Bob and Eve 1 but should remain hidden from Eve 2.
The confidential message $m_2$ is decoded by Bob, while remaining secret from both Eve 1 and  Eve 2. 
Since $m_0$ is not secret, we say that it belongs to layer $0$. 
Similarly, $m_1$ and $m_2$ are  referred to as the layer-1, and layer-2 messages, respectively.
It is assumed that the broadcast channel is degraded. That is, there exist degrading channels 
$\Dset_{B\rightarrow E_1}$ and $\Gset_{E_1\rightarrow E_2}$ such that
\begin{align}
    \channel_{A\rightarrow E_1}=\Dset_{B\rightarrow E_1}\circ\channel_{A\rightarrow B}
    \intertext{and}
     \channel_{A\rightarrow E_2}=\Gset_{E_1\rightarrow E_2}\circ\channel_{A\rightarrow E_1}
\end{align}
where $\channel_{A\rightarrow B}$ and $\channel_{A\rightarrow E_j}$, $j=1,2$, are the marginal channels of the quantum broadcast channel $\channel_{A\rightarrow BE_1 E_2}$.

\vspace{-0.75cm}
\begin{center}
\begin{figure}[htb]
\includegraphics[scale=0.8,trim={2.5cm 10.5cm 0 10cm},clip]
{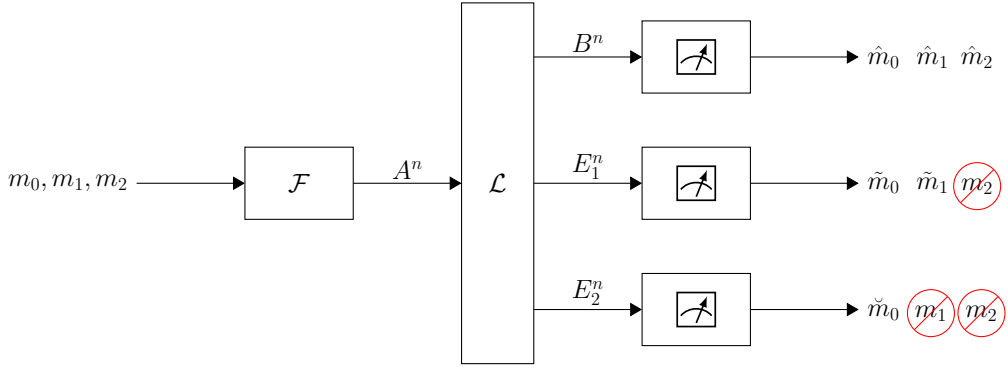} %
\caption{The quantum  broadcast channel with layered secrecy.
}
\label{fig:layeredSec}
\end{figure}
\end{center}

\subsection{Layered-Secrecy Coding}
\label{subsec:McodingLS}
We define a layered-secrecy code, where  a common message  is sent to all receivers,  at a rate $R_0$, and two confidential messages  are sent at rates $R_1$ and $R_2$

\begin{definition} %
\label{def:ClcapacityLS}
A $(2^{nR_0},2^{nR_1},2^{nR_2},n)$ layered-secrecy  code for the quantum broadcast channel $\channel_{A\rightarrow B E_1 E_2}$  consists of the following:
three index sets  $[1:2^{nR_j}]$, for $j=0,1,2$, corresponding to the common message for all users, the confidential message of User 1 and User 2, and the confidential message for User 1 alone, respectively;
 an encoding map $\Fset_{ A^n}$ from the product set $[1:2^{nR_0}]\times [1:2^{nR_1}]\times [1:2^{nR_2}]$ to the input Hilbert space $\Hset_{A^n}$; and	
three decoding POVMs,   $\{\Gamma^{m_0,m_1,m_2}_{B^n}\}  $ for Bob and $\{\Xi^{m_0,m_1}_{E_1^n}\}  $, $\{\Upsilon^{m_0}_{E_2^n}\}  $  for Eve 1 and Eve 2, respectively. %
We denote the code by $(\Fset,\Gamma,\Xi,\Upsilon)$.

The communication scheme is depicted in Figure~\ref{fig:layeredSec}. 
The sender Alice has the system  $A^n$, and the receivers Bob, Eve 1,  and Eve 2 have the systems $B^n$, $E_1^n$, and $E_2^n$,  respectively.
Alice chooses a common message $m_0\in [1:2^{nR_0}]$ that is intended for all users, a layer-1 confidential message $m_1\in [1:2^{nR_1}]$ for Bob and Eve 1,  and a layer-2 confidential message $m_2\in [1:2^{nR_2}]$,
all drawn uniformly at random.
She encodes the messages by applying the encoding map $\Fset_{  A^n}$   which results in an input state
$%
\rho^{m_0,m_1,m_2}_{A^n}= \Fset_{ A^n}( m_0,m_1,m_2  ) %
$, %
 and transmits the system $A^n$ over $n$ channel uses of $\channel_{A\rightarrow B E_1 E_2}$. Hence, the output state is
\begin{align}
\rho^{m_0,m_1,m_2}_{ B^n E_1^n E_2^n}=\channel^{\otimes n} (\rho^{m_0,m_1,m_2}_{A^n}) \,.
\end{align}
Eve 2 receives the channel output system $E_2^n$, and performs a measurement with the POVM 
$\{\Upsilon^{m_0}_{ E_2^n}\}  $. From the measurement outcome, she obtains an estimate of the common message $\breve{m}_0\in [1:2^{nR_0}]$.
 Similarly, Eve 1 finds an estimate of the message pair $(\tm_0,\tm_1)\in [1:2^{nR_0}]\times [1:2^{nR_1}]$ by performing a POVM $\{\Xi^{m_0,m_1}_{E_2^n}\}  $ on the output system $E_1^n$.
 Bob  estimates all three messages by applying $\{\Gamma^{m_0,m_1,m_2}_{B^n}\}  $ on  $B^n$.

The performance of the layered-secrecy code is measured in terms of  the probability of decoding error and the amount of confidential information that is leaked to the non-intended receivers.
The conditional probability of error of the code,   
given that the message pair $(m_0,m_1,m_2)$ was sent, is given by 
\begin{align}
P_{e|m_0,m_1,m_2}^{(n)}(\Fset,\Gamma,\Xi,\Upsilon)&= 
1-\trace[  (\Gamma^{m_0,m_1,m_2}_{B^n}\otimes  \Xi^{m_0,m_1}_{E_1^n}\otimes  \Upsilon^{m_0}_{E_2^n})  \rho^{m_0,m_1,m_2}_{ B^n E_1^n E_2^n} ] \,.
\end{align}
The layer-1 confidential message $m_1$ needs to remain secret from Eve 2, and the layer-2 confidential message $m_2$ needs to remain secret from both Eve 1 and Eve 2. Thereby, 
the  leakage rates of the code $(\Fset,\Gamma,\Xi,\Upsilon)$ are defined as
\begin{align}
s_1^{(n)}(\Fset)&\triangleq   I(M_1;E_2^n|M_0)_\rho
\\
s_2^{(n)}(\Fset)&\triangleq   I(M_2;E_1^n, E_2^n|M_0)_\rho
\label{eq:leakageLS}
\end{align}
where $M_j$ is a classical random variable that is uniformly distributed over the corresponding message index set, $[1:2^{nR_j}]$, for $j=1,2$.

A $(2^{nR_0},2^{nR_1},2^{nR_2},n,\eps,\delta)$ confidential code satisfies 
$%
\frac{1}{2^{n(R_0+R_1)}}\sum_{m_0,m_1} P_{e|m_0,m_1}^{(n)}(\Fset,\Gamma,\Xi)\leq\eps $ %
and $s^{(n)}_j(\Fset)\leq \delta$ for   $j=0,1$.  %
A rate tuple $(R_0,R_1,R_2)$, where $R_j\geq 0$, $j=0,1,2$, is  achievable if for every $\eps,\delta>0$ and sufficiently large $n$, there exists a 
$(2^{nR_0},2^{nR_1},2^{nR_2},n,\eps,\delta)$ layered-secrecy code. 
 The operational layered-secrecy capacity region $\opC_{\text{LS}}(\channel)$ of the quantum broadcast channel $\channel_{A\rightarrow BE_1 E_2}$
is defined as the set of achievable tuples $(R_0,R_1,R_2)$ with layered secrecy. 

As before,
in the bosonic case, it is assumed that the encoder uses a coherent state protocol with an input constraint. Specifically, the input state is a coherent state $\ket{f(m_0,m_1,m_2)}$, where the encoding function, $f: [1:2^{nR_0}]\times [1:2^{nR_1}]\times [1:2^{nR_2}]\to \mathbb{C}^n$,
satisfies
$\frac{1}{n}\sum_{i=1}^n |f_{i}(m_0,m_1,m_2)|^2\leq N_A$.
\end{definition}

\begin{remark}
Here, the index $j$ of the rate $R_j$ corresponds to the secrecy layer. The common message rate $R_0$ is associated with non-secret information, \ie layer $0$.
The confidential message rate $R_1$ 
is associated with layer-1 information, which is secret from one of the users, namely, Eve 2. 
In the WiFi network example mentioned above, this means that Bob and Eve 1 have a higher clearance to access classified information. The confidential message rate $R_2$ 
is associated with a highly classified layer-2 information, and Bob is the only one who has the authority to access this information. As presented by Zou \etal \cite{ZouLiangLaiPoorShamai:15p}, the model can be further generalized to $N$ receivers with $N$ secrecy layers, where User $j$ should only decode the messages $m_0,\ldots,m_{N-j}$, which belong to the $j^{\text{th}}$ layer.
\end{remark}

\begin{remark}
\label{rem:wiretapLS}
Removing Eve 1 (say, $E_1$ has a single dimension), the model reduces to the quantum broadcast  channel with confidential messages, which in turn generalizes  the wiretap channel (see Remark~\ref{rem:wiretap}).
In the case of the broadcast channel with confidential messages, we only have layer-0 and layer-2 messages, whereas the layer-1 rate is $R_1=0$.
\end{remark}

\begin{remark}
As pointed out in Remark~\ref{rem:Qcapacity}, if one considers the transmission of quantum states, instead of classical information, then secrecy is guaranteed by default by the no-cloning theorem.
As the quantum ``message" state cannot be recovered by more than one receiver, this means that the layer-0 and layer-1 quantum rates (in units of qubits per channel use)  must be zero, and we can only have layer-2 communication.
\label{rem:QcapacityLS}
\end{remark}

\subsection{Main Results -- Finite Dimensions}
Consider the quantum degraded broadcast channel $\channel_{A\rightarrow B E_1 E_2}$ with layered secrecy.
We obtain a regularized formula for the capacity region in this setting. Define
\begin{align}
\mathcal{R}_{\text{LS}}(\channel)=
\bigcup_{ p_{X_0,X_1,X_2} \,,\; \varphi_{A}^{x_0,x_1,x_2} } \left\{
\begin{array}{lrl}
(R_0,R_1,R_2)\,:\; &R_0\leq&  I(X_0;E_2)_\rho\\
               &R_1\leq&  [I(X_1;E_1|X_0)_\rho-I(X_1;E_2|X_0)_\rho]_{+}\\
                &R_2\leq&   [I(X_2;B|X_0,X_1)_\rho-I(X_2;E_1 E_2|X_0,X_1)_\rho]_{+}  
\end{array} \right\}\,,
\label{eq:inCls}
\end{align}
where the union is over the distribution of the auxiliary random variables $X_0$, $X_1$, $X_2$ and the collections of quantum states $\{\varphi_{A}^{x_0,x_1,x_2}\}$, with
\begin{align}
    \rho_{X_0 X_1 X_2 B E_1 E_2}=\sum_{x_0,x_1,x_2} p_{X_0,X_1,X_2}(x_0,x_1,x_2)\channel_{A\rightarrow B E_1 E_2}(\varphi_{A}^{x_0,x_1,x_2})\,.
\end{align}

\begin{theorem}
\label{theo:LS}
The layered-secrecy capacity region of the quantum degraded broadcast channel $\channel_{A\rightarrow B E_1 E_2}$ in finite dimensions is given by 
\begin{align}
\opC_{\text{LS}}(\channel)= \bigcup_{n=1}^\infty \frac{1}{n} \mathcal{R}_{\text{LS}}(\channel^{\otimes n}) \,.
\label{eq:Cls}
\end{align}
\end{theorem}
The proof of Theorem~\ref{theo:LS} is given in Appendix~\ref{app:LS}.
\begin{remark}
 Given a classical degraded broadcast channel $W_{Y Z_1 Z_2|X}$, the bound on the top-secret rate $R_2$ in (\ref{eq:inCls}) can be simplified. 
 In particular, we can choose $X_2$ to be the channel input, \ie $X_2=X$. Furthermore, the auxiliary random variables can be chosen such that  $X_0\Cbar X_1\Cbar X\Cbar Y\Cbar Z_1\Cbar Z_2$ form a Markov chain, given a physically degraded broadcast channel. That is, the joint distribution of those variables can be expressed as
 \begin{align}
     p_{X_0 X_1  X Y Z_1 Z_2 }(x_0,x_1,x,y,z_1,z_2)=
     p_{X_0}(x_0) p_{X_1|X_0}(x_1|x_0) p_{X|X_1}(x|x_1) W_{Y|X}(y|x) D_1(z_1|y)D_2(z_2|z_1)\,,
 \end{align}
 for some input distributions $p_{X_0}$, $p_{X_1|X_0}$, and $p_{X|X_1}$, and degrading channels $D_j$, for $j=1,2$. In this case, we have
 \begin{align}
     I(X;Y|X_0,X_1)-I(X;Z_1 Z_2|X_0,X_1)=
     I(X;Y|X_1)-I(X;Z_1|X_1) \,.
 \end{align}
 The last equation also holds for a stochastically-degraded classical channel %
 \cite{ElGamalKim:11b}.
 On the other hand, in the quantum case, the identity $H(B|E_1 E_2)_\rho=H(B|E_1)_\rho$ holds if only if $B\Cbar E_1\Cbar E_2$ form a quantum Markov chain, which is not guaranteed in our model. %
 A quantum Markov chain is defined as follows. The quantum systems $A\Cbar B\Cbar C$ form a Markov chain if and only if there exists a recovery channel $\mathscr{R}_{B\rightarrow BC}$ such that 
 $\rho_{ABC}=(\mathrm{id}_A\otimes \mathscr{R}_{B\rightarrow BC})(\rho_{AB})$ \cite{Sutter:18b}.
However, as pointed out in \cite{YardHaydenDevetak:11p}, the fact that the quantum broadcast channel is degraded does not imply that  $B\Cbar E_1\Cbar E_2$ form a quantum Markov chain. 
 \end{remark}

\begin{center}
\begin{figure}[tb]
\includegraphics[scale=0.85,trim={1cm 10.5cm 0 10cm},clip]
{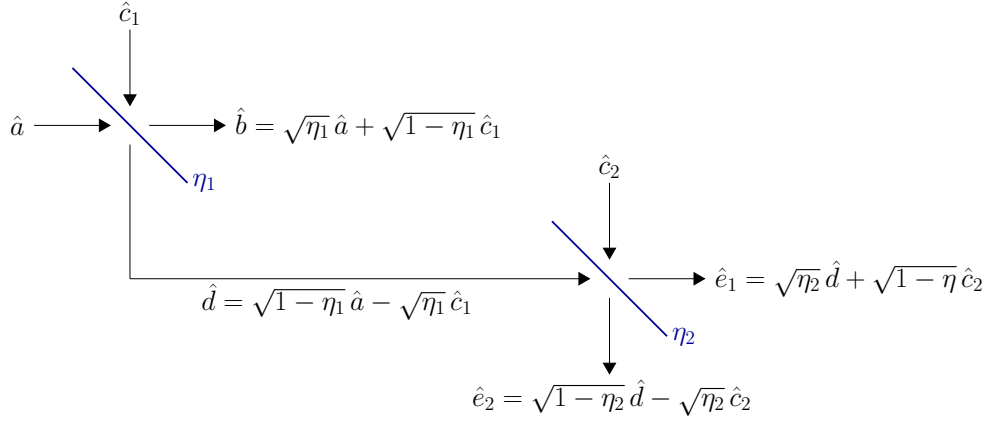} %
\caption{The 3-receiver bosonic broadcast channel with layered secrecy.
}
\label{fig:BSpLayeredS}
\end{figure}
\end{center}

\subsection{Main Results -- Bosonic Channel}
In this section, we consider layered secrecy for the pure-loss bosonic broadcast channel, specified by 
\begin{align}
\hb&=\sqrt{\eta_1}\, \ha +\sqrt{1-\eta_1}\,\hc_1 \\
\hd&=\sqrt{1-\eta_1}\, \ha -\sqrt{\eta_1}\,\hc_1
\intertext{and}
\he_1&=\sqrt{\eta_2}\, \hd +\sqrt{1-\eta_2}\,\hc_2 \\
\he_2&=\sqrt{1-\eta_2}\, \hd -\sqrt{\eta_2}\,\hc_2
\end{align}
where $\hc_j$ is associated with the environment noise and the parameter $\eta_j$  is %
the transmissivity, $\frac{1}{2}\leq \eta_j\leq 1$, which depends on the length of the optical fiber and its absorption length, for $j=1,2$. The relation above corresponds to a concatenation of two beam splitters. It is assumed that the noise modes $\hc_1$ and $\hc_2$ are uncorrelated and in the vacuum state. As illustrated in Figure~\ref{fig:BSpLayeredS}, 
the 3-receiver pure-loss bosonic broadcast channel corresponds to the operation of two  beam splitters. 

We determine the capacity region. Denote the channel by $\channel_{\text{pure-loss}}$.

 Our result relies on the long-standing strong minimum output-entropy conjecture, Conjecture~\ref{con:MinOutStrong}. %

\begin{theorem}
\label{theo:PlossCfLS}
Assume that Conjecture~\ref{con:MinOutStrong} holds.
Then, the layered-secrecy capacity region of the pure-loss bosonic broadcast channel  is bounded by %
\begin{align}
    \opC_{\text{LS}}(\channel_{\text{pure-loss}})\supseteq
    \bigcup_{ 
    \substack{ \beta_1,\beta_2\geq 0 \,,\\
    \beta_1+\beta_2\leq 1}} \left\{
\begin{array}{rl}
(R_0,R_1)\,:\; R_0\leq&  g\big((1-\eta_1)(1-\eta_2) N_A\big)-g\big((\beta_1+\beta_2)(1-\eta_1)(1-\eta_2)  N_A\big) \\
							 R_1\leq& g\big((\beta_1+\beta_2)\eta_2  (1-\eta_1) N_A\big)-g\big(\beta_2 \eta_2  (1-\eta_1) N_A\big)\\
				&-[g\big((\beta_1+\beta_2)(1-\eta_2)  (1-\eta_1) N_A\big)-g(\beta_2(1-\eta_2)  (1-\eta_1) N_A\big)]
				\\			 R_2\leq& g\big(\eta_1 \beta_2 N_A\big)-g\big((1-\eta_1) \beta_2 N_A\big)
\end{array} \right\} \,.
\end{align}
\end{theorem}
To show %
Theorem~\ref{theo:PlossCfLS},
we extend the finite-dimension result in Theorem~\ref{theo:LS} to the bosonic channel with infinite-dimension Hilbert spaces based on the discretization limiting argument by Guha \etal \cite{GuhaShapiroErkmen:07p} (see also Subsection~\ref{subsec:Discretization} for further expalanation). 
If one ignores the input constraint, then based on Theorem~\ref{theo:LS}, the region $\mathcal{R}_{\text{LS}}(\channel_{\text{pure-loss}})$ is achievable, \ie
$\opC_{\text{LS}}(\channel_{\text{lossy}})\supseteq \mathcal{R}_{\text{LS}}(\channel_{\text{lossy}})$.
Thus, we obtain the following inner bound,
\begin{align}
    \opC_{\text{LS}}(\channel_{\text{lossy}})\supseteq   \left\{
 \begin{array}{rl}
 (R_0,R_1)\,:\; R_0\leq& I(X_0;E_2)_\rho\\
 							 R_1\leq&  I(X_1;E_1|X_0)_\rho-I(X_1;E_2|X_0)_\rho \\
 							 R_2\leq&  I(X_2;B|X_0,X_1)_\rho-I(X_2;E_1 E_2|X_0,X_1)_\rho 
 \end{array} \right\}\,,
\end{align}
where the auxiliary variables $X_j$  can be chosen arbitrarily. Given the input constraint, we need to add the restriction  
$\mathbb{E}(|X_2|^2)\leq N_A$.

Then, set the input to be a coherent state with $X_1=X_0+Q_1$ and $X_2=X_1+Q_2$, where
 $X_j$ and $Q_j$ are  independent complex Gaussian random variables, 
\begin{align}
X_0 &\sim \mathcal{N}_{\mathbb{C}}(0,(1-\beta_1-\beta_2)\frac{N_A}{2})\\
Q_j &\sim \mathcal{N}_{\mathbb{C}}(0,\beta_j\frac{N_A}{2})
\end{align}
for some $\beta_j\in [0,1]$, $j=1,2$.
Then, $X_1$ and $X_2$ are distributed according to
\begin{align}
X_1&\sim \mathcal{N}_{\mathbb{C}}(0,\frac{(1-\beta_2)N_A}{2})\\
X_2&\sim \mathcal{N}_{\mathbb{C}}(0,\frac{N_A}{2})\,.
\end{align}

Consider the first beam splitter.
Since the conditional distribution of $X_2$ given $X_1=x_1$, $X_0=x_0$ is $\sim\mathcal{N}_{\mathbb{C}}(0,\beta_2\frac{N_A}{2})$,
the mutual informations corresponding to each output are given by
\begin{align}
    I(X_2;B|X_0,X_1)&=H(B|X_0,X_1)_\rho=
    g(\eta_1\beta_2 N_A)\\
    I(X_2;E_1 E_2|X_0,X_1)&=H(D|X_0,X_1)=g\big((1-\eta_1)\beta_2 N_A\big)
    \label{eq:Eve12entDir}
\end{align}
where the first equality in (\ref{eq:Eve12entDir}) holds since the broadcast channel from $D$ to $E_1 E_2$ is isometric.

As for the second beam splitter, observe that the mean photon number of the input is $N_D=(1-\eta_1)N_A$. Hence, the entropies corresponding to each output are given by
\begin{align}
    H(E_1)_\rho&=g(\eta_2 N_D)=g\big(\eta_2 (1-\eta_1) N_A\big)\\
    H(E_1|X_0)_\rho&=g\big(\eta_2 (\beta_1+\beta_2) N_D\big)=g\big(\eta_2 (\beta_1+\beta_2) (1-\eta_1) N_A\big)\\
    H(E_1|X_0,X_1)_\rho&=g(\eta_2 \beta_2 N_D)=g\big(\eta_2 \beta_2 (1-\eta_1) N_A\big)%
    \intertext{and}
    H(E_2)_\rho&=g\big( (1-\eta_2) N_D\big)=g\big( (1-\eta_2) (1-\eta_1)N_A\big)\\
    H(E_2|X_0)_\rho&=g\big((1-\eta_2) (\beta_1+\beta_2) N_D\big)=g\big((1-\eta_2) (\beta_1+\beta_2) (1-\eta_1)N_A\big)\\
    H(E_2|X_0,X_1)_\rho&=g\big((1-\eta_2) \beta_2 N_D\big)=g\big((1-\eta_2) \beta_2 (1-\eta_1)N_A\big)%
\end{align}
Thus,
\begin{align}
    I(X_0;E_2)_\rho&=H(E_2)_\rho-H(E_2|X_0)_\rho=g\big( (1-\eta_2) (1-\eta_1)N_A\big)-g\big((1-\eta_2) (\beta_1+\beta_2) (1-\eta_1)N_A\big)
    \\
    I(X_1;E_1|X_0)_\rho&=H(E_1|X_0)_\rho-H(E_1|X_0,X_1)_\rho=g\big(\eta_2 (\beta_1+\beta_2) (1-\eta_1) N_A\big)-g\big(\eta_2 \beta_2 (1-\eta_1) N_A\big)\\
    I(X_1;E_2|X_0)_\rho&=H(E_1|X_0)_\rho-H(E_1|X_0,X_1)_\rho=g\big((1-\eta_2) (\beta_1+\beta_2) (1-\eta_1) N_A\big)-g\big((1-\eta_2) \beta_2 (1-\eta_1) N_A\big)
\end{align}
This completes the achievability proof.

\section{Discussion}
\subsection{Discretization}
\label{subsec:Discretization}
Many capacity theorems for Gaussian channels,  in both classical and quantum information theory, are derived by extending the finite-dimension results to the continuous infinite-dimension Gaussian channel \cite{ElGamalKim:11b}.
This requires a discretization limiting argument, as \eg in \cite{GuhaShapiroErkmen:07p}. 
This approach is often more convenient than devising a coding scheme and perform the analysis ``from scratch". Yet, such a proof is less transparent and may give less insight for the design of practical error-correction codes (see also discussion in \cite{PeregSteinberg:21p}).

The most common discretization approach is
based on the following operational argument. Consider a classical memoryless channel $W_{Y|X}$, with  continuous input $X$ and output $Y$.
We can construct a codebook while restricting ourselves to discrete values, in $\{-L\delta,-(L-1)\delta,\ldots,-\delta,0,\delta,\ldots, (L-1)\delta,L\delta \}$, and the decoder can also discretize the received signal, with an arbitrarily small discretization step $\delta>0$ and arbitrarily large $L>0$. In this manner, we are effectively coding over a finite-dimension channel, with input $X_{\delta}$ and ouput $Y_{\delta}$ over finite alphabets.
Thus, by the finite-dimension capacity result, for every input distribution $p_{X_{\delta}}$, a rate $R<I(X_{\delta};Y_{\delta})-\eps$ is achievable, for arbitrarily small $\eps>0$.
The achievability proof can then be completed by analyzing the limit of the mutual information $I(X_{\delta};Y_{\delta})$ as $\delta$ tends to zero.
A similar argument can be applied for the bosonic channel, where the input is restricted to coherent states of discretized values and the output dimension is restricted by the decoding measurement.
For basic channel networks, the Gaussian capacity result can be obtained directly. However, in adversarial models, such as the wiretap channel, this approach may become tricky. In particular, a straightforward application will force the eavesdropper to discretize her signal. Clearly, this does not make sense operationally.

An alternative discretization approach, which can be applied to adversarial models as well, is based on continuity arguments.
In particular, we can view the operational capacity $C(W)$ as an unknown functional of the probability measure $W_{Y|X}$, which may have either finite or infinite dimensions. 
Then, if $W_{Y|X}$ is a Gaussian channel, then we can define a sequence of discretized channels $\{ \bar{W}^{\delta_k}_{Y|X} \}_{k\geq 1}$, where $\delta_1,\delta_2,\ldots$ converges to zero uniformly. 
Since the Gaussian distribution is continuous and smooth, the sequence of probability measures $\{ \bar{W}^{\delta_k}_{Y|X} \}_{k\geq 1}$ converges to the Gaussian probability measure $W_{Y|X}$. Furthermore, we have 
\begin{align}
    \lim_{\delta\to 0} \bar{W}^{\delta}_{Y|X}=W_{Y|X} \,.
    \label{eq:ChContinuity}
\end{align}
This follows from the observation  that in the Gaussian case, the discretized measure $W^{\delta}_{Y|X}$ is continuous in $\delta$.
Now, given a channel $\bar{W}_{Y|X}$ in finite dimensions, let $\inR(\bar{W})$ denote the capacity formula.
Then, based on the finite-dimension capacity theorem, the capacity formula $\inR(\bar{W})$ of a channel $\bar{W}$ in finite dimensions is continuous in the channel parameters, since the mutual information $I(X;Y)$ is a continuous functional in the joint distribution $p_{X,Y}$, which can be extended to  quantum channels as well \cite{Shirokov:17p}. The finite-dimension capacity theorem states that the operational capacity and the capacity formula are identical, \ie $C(\bar{W})=\inR(\bar{W})$, for a finite-dimension channel $\bar{W}$. This, in turn, implies that the operational capacity  $C(\bar{W})$ is also a continuous functional.

In general, a composite of two continuous  functions $f:\Aset\to\Bset$ and $g:\Bset\to\mathbb{R}$ satisfies 
$\lim_{x\rightarrow x_0} f(g(x))=f(\lim_{x\rightarrow x_0} g(x))$, for every $x_0\in\Aset$ such that those limits exist, and this property can also be extended to limits of sequences. Thereby, the limit of the capacities of the discretized channels converges to the capacity of the Gaussian channel. Specifically,
\begin{align}
    C(W)&\stackrel{(a)}{=} C\left(\lim_{\delta\rightarrow 0}\bar{W}^{\delta} \right)
    \nonumber\\
    &\stackrel{(b)}{=}\lim_{\delta\rightarrow 0} C\left(\bar{W}^{\delta}\right)
    \nonumber\\
    &
    \stackrel{(c)}{=}\lim_{\delta\rightarrow 0}\inR\left(\bar{W}^{\delta}\right)
        \nonumber\\
    &
    \stackrel{(d)}{=}\inR\left(\lim_{\delta\rightarrow 0}\bar{W}^{\delta}\right)
        \nonumber\\
    &\stackrel{(a)}{=} \inR(W)
    \,,
\end{align}
where $(a)$ follows from  the convergence property in  (\ref{eq:ChContinuity}), $(b)$ holds by the continuity of the operational capacity and the  Gaussian channel measure, $(c)$ follows from the finite-dimension capacity theorem, and $(d)$ from the continuity of the  mutual-information  formula and the  Gaussian channel measure.
We note that we have repeatedly used the continuity of the channel distribution and the convergence of the discretized sequence, for a discretization of our choice. Thus, the arguments above are not suitable for a general channel measure. The advantage of this discretization approach, using continuity arguments, is that it yields the capacity of Gaussian and bosonic channels in a direct manner, even with security requirements.
As before, a similar argument can be applied for the bosonic channel, where the input is now restricted to discretized-value coherent states   and the output dimension is restricted by the decoding measurement.

The disadvantage of this approach is that it removes the operational meaning of the capacity and error-correction codes from consideration, and turns the problem into a calculus exercise. This makes it difficult to gain insight on the design of coding techniques for Gaussian channels.

\subsection{Strong Minimum Output-Entropy Conjecture}
The converse part of our result on the pure-loss bosonic broadcast channel with confidential messages relies on the strong minimum output-entropy conjecture.
In the single-user case, the conjecture is not required.
As previously mentioned, the single-user wiretap channel can be obtained from the broadcast model with confidential messages by restricting the common message rate to $R_0=0$ (see Remark~\ref{rem:wiretap}). That is, in the wiretap setting, Eve eavesdrops on the confidential message of Bob, and she is not required to decode any messages. 
In the special case of an isometric wiretap channel $\Uset_{A\to BE}$, Eve's system is interpreted as Bob's environment, and the secrecy capacity of the wiretap channel is referred to as the private capacity of the main channel $\channel_{A\rightarrow B}$ (see Remark~\ref{rem:privateC}).
Denote the private capacity by $C_{\text{P}}(\channel)$.
For a degradable isometric channel in finite dimension, the private capacity equals the quantum capacity, and it is given by
\begin{align}
    C_{\text{P}}(\channel)= \max_{ \ket{\phi_{AA'}}}[H(B)_\rho-H(E)_\rho]
\end{align}
with $\rho_{A'BE}\equiv (\Uset_{A\to BE}\otimes \mathrm{id})(\phi_{AA'}) $ (see \cite[Theorem 13.6.2]{Wilde:17b}). As Wilde and Qi established in \cite{WildeQi:18p}, in the bosonic case under input constraint $N_A$, the maximum is achieved for a thermal state $\tau(N_A)$ (see Theorem 6 therein). Hence, the private capacity of the pure-loss bosonic channel is given by
\begin{align}
    C_{P}(\channel_{\text{pure-loss}})= [g(\eta N_A)-g((1-\eta)N_A)]_{+} \,.
\end{align}
Wilde and Qi's  derivation \cite{WildeQi:18p} for this property is based on the following argument. Based on the monotonicity of the divergence with respect to quantum channels,  we have $D(\rho_B || \tau(\eta N_A))\geq D(\rho_E || \tau((1-\eta) N_A)) $, as Eve's channel is degraded with respect to Bob's channel. This, in turn, can be written as
\begin{align}
   H(B)_\rho-H(E)_\rho\leq \trace(\rho_E \log \tau((1-\eta) N_A))-\trace(\rho_B \log \tau(\eta N_A)) \,.
\end{align}
Plugging a thermal state, $\tau(N)=\frac{e^{-NG}}{\trace(e^{-NG})}$ where $G$ is a Gibbs observable, into the RHS, we obtain
\begin{align}
    \trace(\rho_E \log \tau((1-\eta) N_A))-\trace(\rho_B \log \tau(\eta N_A))&=\log\frac{\trace(e^{- N_A G})}{\trace(e^{-(1-\eta)N_A G})}
    -\trace(G\rho_E)  +\trace(G\rho_B) 
    \nonumber\\
    &\leq 
    \log\frac{\trace(e^{- N_A G})}{\trace(e^{-(1-\eta)N_A G})}
    -(1-\eta)N_A  +\eta N_A 
\end{align}
since $\trace(G\rho_B) \leq \eta N_A$ and $\trace(G\rho_E) \geq \eta N_A$.
The proof follows since the inequalities above are saturated for a thermal input state $\rho_A=\tau(N_A)$. Unfortunately, this technique does not seem to yield the desired result for a broadcast channel.

The minimum output-entropy conjecture, Conjecture~\ref{con:MinOutStrong}, is known to hold in special cases \cite[Remark 2]{DePalma:19p}. In particular, the conjecture was proved
to hold for $n=1$. 
This weak version of the minimum output-entropy property was established by De Palma \etal \cite{DePalmaTrevisanGiovannetti:17p}, in 2017,  using Lagrange multiplier techniques (see also \cite{QiWildeGuha:16a}).
However, as pointed out in \cite[Sec. V]{DePalmaTrevisanGiovannetti:17p}, this is insufficient for the converse proof of the bosonic broadcast channel, which requires the strong minimum output-entropy conjecture.

\section{Acknowledgments}
Uzi Pereg and Roberto Ferrara were supported by the German Bundesministerium f\"ur Bildung und Forschung (BMBF) through Grant n. 16KIS0856. Matthieu Bloch was supported by the American National Science Foundation (NSF) through Grant n. 1955401. Pereg was also supported by the Israel CHE Fellowship for Quantum Science and Technology.

 \begin{appendices}
 \section{Proof of Lemma~\ref{lemm:PureS}}
 \label{app:PureS}
 Consider the region $\inC_{\text{k-a}}(\channel)$ as defined in (\ref{eq:inCskG}).

\subsection{Cardinality Bounds}
To bound the alphabet size of the random variables $T$ and $X$, we use the Fenchel-Eggleston-Carath\'eodory lemma \cite{Eggleston:66p} and arguments similar to \cite{YardHaydenDevetak:08p}.
Let
\begin{align}
N_0=&|\Hset_A|^4+3 \\
N_1=& |\Tset|(|\Hset_A|^4+2) \,.
\end{align}

First, fix $p_{X|T}(x|t)$, and consider the ensemble $\{ p_T(t)p_{X|T}(x|t) \,, \varphi_A^{t,x}  \}$. 
Every mixed state $\varphi_A$ has a unique parametric representation $u(\varphi_A)$ of dimension $|\Hset_A|^4-1$, since  the corresponding density matrix has $|\Hset_A|^2$ complex entries and the constraint $\trace(\varphi_A)=1$. 
Then, define a map $f_0:\Xset\rightarrow \mathbb{R}^{N_0}$ by
\begin{align}
f_0(t)= \left(  u(\rho_A^t) \,,\; H(B|T=t)_\rho \,,\; H(E|T=t)_\rho \,,\;
H(B|X,T=t)_\rho \,,\; H(E|X,T=t)_\rho 
\right)
\end{align}
where $\rho_A^t=\sum_{x} p_{X|T}(x|t)  \theta_A^{t,x}$. The map $f_0$ can be extended to probability distributions as follows,
\begin{align}
F_0 \,:\; p_T  \mapsto
\sum_{t\in\Tset} p_T(t) f_0(t)= \left(  u(\rho_A) \,,\; H(B|T)_\rho \,,\; H(E|T)_\rho \,,\;
H(B|X,T)_\rho \,,\; H(E|X,T)_\rho     \right) %
\end{align}
where $\rho_A=\sum_t p_T(t) \rho_A^t$.
According to the Fenchel-Eggleston-Carath\'eodory lemma \cite{Eggleston:66p}, any point in the convex closure of a connected compact set within $\mathbb{R}^d$ belongs to the convex hull of $d$ points in the set. 
Since the map $F_0$ is linear, it maps  the set of distributions on $\Tset$ to a connected compact set in $\mathbb{R}^{N_0}$. Thus, for every  $p_T$, 
there exists a probability distribution $p_{\bar{T}}$ on a subset $\overline{\Tset}\subseteq \Tset$ of size $%
N_0$, such that 
$%
F_0(p_{\bar{T}})=F_0(p_{T}) %
$. %
We deduce that the alphabet size can be restricted to $|\Tset|\leq N_0$, while preserving $\rho_A$ and
$\rho_{BE}\equiv \channel_{A\rightarrow BE}(\rho_A)$; $I(T;B)_\rho=H(B)_\rho-H(B|T)_{\rho}$,
$I(X;B|T)_\rho=H(B|T)_\rho-H(B|X,T)_{\rho}$;
 and similarly, $I(T;E)_\rho$ and $I(X;E|T)_\rho$.

We move to the alphabet size of $X$. We keep  $\Tset$ and $p_T$ fixed. Then,  define the map $f_1^{(t)}:\Xset\rightarrow \mathbb{R}^{N_1}$ by
\begin{align}
f^{(t)}_{1}(x)=& \left(     p_{T}(t) \,,\; u(\varphi_A^{t,x})   \,,\; H(B|X=x,T=t)_{\rho} \,,\; H(E|X=x,T=t)_{\rho}  \right) %
\end{align}
for $t\in\Tset$.
 Now, the extended map is
\begin{align}
F^{(t)}_1 \,:\; p_{X|T=t} \mapsto  \sum_{x\in\Xset} p_{X|T}(x|t) f^{(t)}_1(x) = \left( p_{T}(t)\,,\; u(\rho_A^{t})  \,,\; H(B|X,T=t)_{\rho} \,,\; H(E|X,T=t)_{\rho}  \}_{t\in\Tset}   \right) \,.
\end{align}
By the Fenchel-Eggleston-Carath\'eodory lemma \cite{Eggleston:66p}, for every  $p_{X|T}$, 
there exists $p_{\bar{X}|T}$ on a subset $\overline{\Xset}\subseteq \Xset$ of size $ N_1 $, such that 
$F_1^{(t)}(p_{\bar{X}|T=t})=F_1^{(t)}(p_{X|T=t})$ for all $t\in\Tset$.
Thus, we can restrict the alphabet size  to $|\Xset|\leq N_1 $, while preserving  $\rho_A^t$, $H(B|T)_\rho=
\sum_{t} H(B|T)_{\channel^{(1)}(\sigma_A^t)}$ and similarly $H(B|T)_\rho$; $\rho_A$, $\rho_{B,E}$, $I(X;B|T)_\rho$, and $I(X;E|T)_\rho$.

\subsection{Purification}
Suppose that $\channel_{A\rightarrow BE}$ is degraded.
 To prove that a union over pure states is sufficient, we show that for every achievable rate pair $(R_0,R_1)$, there exists a rate pair $(R_0',R_1')$, where $R_j'\geq R_j$ for $j=0,1$,  that can be achieved with pure states.
 Fix $p_{T,X}(t,x)$ and $\{ \varphi^{t,x}_{A} \}$. Let  
 \begin{align}
 R_0\leq& \min\left( I(T;B)_\rho\,,\; I(T;E)_\rho \right) %
 \label{eq:R0p}
 \\
 R_1\leq&  \min\left( I(X;B|T)_\rho- I(X;E|T)_\rho+R_K \,,\;  I(X;B|T)_\rho \right)
 \label{eq:R1p}
 \end{align}
 and consider the spectral decomposition,
 \begin{align}
 \varphi^{t,x}_{A}=\sum_{z\in\Zset} p_{Z|T,X}(z|t,x) \phi^{t,x,z}_A 
 \label{eq:scPureThetaxz}
 \end{align}
 where $p_{Z|T,X}(z|t,x)$ is a conditional  probability distribution, and $\phi^{t,x,z}_A$ are pure. 
 Consider the extended state
 \begin{align}
 \rho_{TXZA}=\sum_{t,x,z} p_{T,X}(t,x)p_{Z|T,X}(z|t,x)
 \ketbra{t} \otimes \ketbra{x} \otimes \ketbra{z} \otimes  \phi^{t,x,z}_A \,.
 \label{eq:scPureSXZWA}
 \end{align}

Now, observe that the union in the RHS of (\ref{eq:inCskG}) includes
 the rate pair $(R_0',R_1')$ that is given by
 \begin{align}
 R_0'&= R_0 %
 \label{eq:scR1p}
 \\
 R_1'&= \min\left( I(X,Z;B|T)_\rho- I(X,Z;E|T)_\rho+R_K \,,\;  I(X,Z;B|T)_\rho \right)
 \label{eq:scD1p}
 \end{align}
 which is obtained by plugging $X'=(X,Z)$ instead of $X$, and the pure states $\phi^{t,(x,z)}_A$ instead of $\varphi_A^{t,x}$.
 That is, $(R_0',R_1')\in\inC_{\text{k-a}}(\channel)$. 
 By the chain rule, 
 \begin{align}
      I(X,Z;B|T)_\rho=I(X;B|T)_\rho+I(Z;B|T,X)_\rho\geq I(X;B|T)_\rho \,.
 \end{align}
 Furthermore,
 $ I(X,Z;B|T)_\rho-I(X,Z;E|T)_\rho=
 [I(X;B|T)_\rho-I(X;E|T)_\rho]
 +[I(Z;B|X,T)_\rho-I(X;E|X,T)_\rho]$. 
 Assuming that the channel is degraded, 
 we have
 $I(Z;B|X,T)_\rho\geq I(Z;E|X,T)_\rho$, by the quantum data processing inequality \cite[Theorem 11.5]{NielsenChuang:02b}.
 Hence,
 \begin{align}
     I(X,Z;B|T)_\rho-I(X,Z;E|T)_\rho\geq 
     I(X;B|T)_\rho-I(X;E|T)_\rho
 \end{align}
 and it follows that $R_1'\geq R_1$.
 Thereby, the union can be restricted to pure states.
\qed

\section{Proof of Theorem~\ref{theo:KAqbc}}
\label{app:KAqbc}
Consider the broadcast channel $\channel_{A\rightarrow BE}$ with confidential messages and key assistance, with a key of rate $R_K$.

\subsection{Achievability proof}
The direct part follows the classical arguments in \cite{Yamamoto:97p,KangLiu:10c}, using rate-splitting to combine the one-time pad coding scheme and the unassisted confidential coding scheme.
We split the confidential message $m_1$ into two parts, one of them is encrypted by the one-time pad encryption, using the key, and the other is encoded without key assistance.

Let  $m_1=(m_{1\mathrm{c}},m_{1\mathrm{k}})$ be a composite message, where $m_{1\mathrm{c}}\in [1:2^{nR_{1\mathrm{c}}}]$ and $m_{1\mathrm{k}}\in [1:2^{nR_{1\mathrm{k}}}]$,  with  $R_{1\mathrm{k}}\leq R_K$, where the subscripts `c' and `k' indicate the confidential and key encodings, respectively.
The overall private rate is 
\begin{align}
    R_1=R_{1\mathrm{k}}+R_{1\mathrm{c}}
\end{align}
Let $k\in [1:2^{n(R_{1\mathrm{k}}-\eps)}]$ be a uniformly distributed key, where $\eps>0$ is an   chosen such that  $n(R_{1\mathrm{k}}-\eps)$ is an integer (as $n$ grows to infinity, we can take $\eps$ to be arbitrarily small).
The bit-wise parity of $m_{1\mathrm{k}}$ and the key can then be represented by %
\begin{align}
    \ell_{1\mathrm{k}}=m_{1\mathrm{k}} +k \,\mod 2^{n(R_{1\mathrm{k}}-\eps)} \,.
\end{align}
We refer to $\ell_{1\mathrm{k}}$ as the encrypted component of the message.

Then, consider a code for the quantum broadcast channel with confidential messages without key assistance, where we transmit a triplet message $(m_0,m_{1\mathrm{c}},\ell_{1\mathrm{k}})$,
where $m_0$ is a common message for both Bob and Eve, $m_{1\mathrm{c}}$ is a private and confidential message of Bob, and $\ell_{1\mathrm{k}}$ is a message of Bob that does not need to satisfy the confidentiality requirement. 
Based on the previous result by Salek \etal \cite[Theorem 3]{SalekHsieFonollosa:19a}, the message triplet $(m_0,m_{1\mathrm{c}},\ell_{1\mathrm{k}})$ can be transmitted with vanishing error probability, $P_{e|m_0,m_{1\mathrm{c}},\ell_1}^{(n)}(\Fset,\Gamma,\Xi)\rightarrow 0$ as $n\rightarrow \infty$, for rate triplets 
$(R_0,R_{1\mathrm{c}},R_{1\mathrm{k}})$ such that
\begin{align}
    R_0&\leq \min\left( I(T;E)_\rho \,,\; I(T;B)_\rho \right)-\eps' \\
    R_{1\mathrm{c}}&< [I(X;B|T)_\rho-I(X;E|T)_\rho-\eps']_{+} \\
    R_{1\mathrm{c}}+R_{1\mathrm{k}}&< I(X;B|T)_\rho-\eps' 
\end{align}
where $\eps'$ is arbitrarily small.
As  $R_1=R_{1\mathrm{c}}+R_{1\mathrm{k}}$,
this reduces to the region in (\ref{eq:inCskG}).
That is, Eve can decode $m_0$ and Bob can decode
$m_0$, $m_{1\mathrm{c}}$, and $\ell_{1\mathrm{k}}$  with vanishing probability of error. Since Bob has the key $k$, he determines the private component $m_{1\mathrm{k}}=\ell_{1\mathrm{k}}+k \,\mod 2^{n(R_{1\mathrm{k}}-\eps)}$.

As for the confidentiality requirement,
the confidential encoding scheme only  guarantees that $M_{1\mathrm{c}}$ is private, \ie Eve's output does not depend on it. Hence,
\begin{align}
    I(M_{1\mathrm{c}};E^n|M_{1\mathrm{k}},M_0)_\rho \leq \delta_n
    \label{eq:Lm1cEnU}
\end{align}
 where $\delta_n\rightarrow 0$ as $n\rightarrow\infty$.
 Since $M_{1\mathrm{k}}$ and $L_{1\mathrm{k}}$ are statistically independent, there is no correlation between the state of Eve's output system $E^n$ and the confidential message $M_{1\mathrm{k}}$ as well, \ie
 \begin{align}
     I(M_{1\mathrm{k}};E^n|M_0)_\rho\leq \delta_n' 
     \label{eq:Lm1kEnU}
 \end{align}
  where $\delta_n'\rightarrow 0$ as $n\rightarrow\infty$.
 Thus, by (\ref{eq:Lm1cEnU}) and (\ref{eq:Lm1kEnU}),
 \begin{align}
     I(M_1;E^n|M_0)_\rho=I(M_{1\mathrm{c}},M_{1\mathrm{k}};E^n|M_0)_\rho\leq \delta_n+\delta_n' \,.
 \end{align}
This completes the achievability proof.

\subsection{Converse proof}
The regularized converse proof is analogous to the classical proof in \cite{SchaeferKhistiPoor:18p}.
 Suppose that Alice and Bob share a uniformly distributed key $K\in [1:2^{nR_K}]$. 
 Alice chooses $M_0$ and $M_1$ uniformly at random. Given $K=k$ and $M_j=m_j$, she prepares an input state $\rho^{m_0,m_1,k}_{ A^n}$. 
The channel output is
$\rho_{B^n E^n}=\frac{1}{2^{n(R_0+R_1+R_K)}}\sum_{m_0,m_1,k} \channel_{A^n\rightarrow B^n E^n}(\rho^{m_0,m_1,k}_{ A^n}) $.
Then, Bob and Eve perform  decoding POVMs $\Gamma^{m_0,m_1}_{B^n|k}$ and $\Xi^{m_0}_{E^n}$, respectively.
Consider a sequence of codes $(\Fset_n,\Gamma_n,\Xi_n)$ with key assistance, such that the average probability of error and the leakage tend to zero, hence
the error probabilities $\prob{ \tM_0\neq M_0 }$, $\prob{ (\hM_0,\hM_1)\neq (M_0,M_1)}$, $\prob{ \hM_1\neq M_1 |M_0}$,   are bounded by some
$\alpha_n$ which tends to zero as $n\rightarrow \infty$.
By Fano's inequality \cite{CoverThomas:06b}, it follows that%
\begin{align}
H(M_0|\tM_0) \leq n\eps_{1n} \,,
\label{eq:Hm0ukEve}
\\
H(M_0|\hM_0) \leq n\eps_{2n}\,,
\label{eq:Hm0ukBob}
\\
H(M_1|\hM_1,M_0) \leq n\eps_{3n}\,,
\label{eq:Hm1uk}
\end{align}
where $\eps_{jn}$ tend to zero as $n\rightarrow\infty$.
Furthermore, the leakage rate is bounded by
\begin{align}
    I(M_1;E^n|M_0)_\rho\leq \delta_n
    \label{eq:leakConU1k}
\end{align}
where $\delta_n$ tends to zero as $n\rightarrow \infty$.

Thus, the common rate is bounded as 
\begin{align}
nR_0&\leq I(M_0;E^n)_{\rho}+n\eps_{1n} 
\nonumber\\
&\leq I(M_0,K;E^n)_{\rho}+n\eps_{1n} 
\label{eq:R0convEk}
\end{align}
by the same arguments as in (\ref{eq:multiUp1a}).
Also,  
\begin{align}
nR_0&= H(M_0)=I(M_0;\tM_0)_{\rho}+H(M_0|\tM_0) 
\nonumber\\
&\leq I(M_0;\tM_0)_{\rho}+n\eps_{2n} \nonumber\\
&\leq I(M_0;B^n, K )_{\rho}+n\eps_{2n} %
\label{eq:multiUp1b}
\end{align}
where the first inequality follows from  (\ref{eq:Hm0ukEve}), and the last inequality %
follows from the Holevo bound. 
Since the key is independent of the messages, 
$I(M_0;K)=0$, and we can re-write the last bound as
\begin{align}
    nR_0&\leq I(M_0;B^n| K )_{\rho}+n\eps_{2n}
    \nonumber\\
    &\leq I(M_0,K;B^n )_{\rho}+n\eps_{2n}
    \label{eq:R0convBk}
\end{align}

Similarly, the private rate is bounded as
\begin{align}
    nR_1 &\leq I(M_1;B^n,K|M_0)_\rho+n\eps_{3n}
    \label{eq:nR1kU1} \\
    &= I(M_1;B^n|M_0,K)_\rho+n\eps_{3n}
    \,. 
    \label{eq:nR1kU2}
\end{align}
As we have seen in (\ref{eq:leakConU1k}) that $-I(M_1;E^n|M_0)_\rho+ n\delta_n\geq 0$ due to the leakage requirement, we deduce that 
\begin{align} 
nR_1&\leq I(M_1;B^n,K|M_0)_\rho-I(M_1;E^n|M_0)_\rho+n\delta_n+n\eps_{3n} \nonumber\\
&= I(M_1;B^n,K|M_0)_\rho-I(M_1;E^n,K|M_0)_\rho+I(M_1;K|E^n,M_0)_\rho+n\delta_n+n\eps_{3n}
\label{eq:nR1upp0k}
\end{align}
Now, since the key is classical, the third term can be bounded by 
\begin{align}
    I(M_1;K|E^n,M_0)_\rho\leq H(K)=nR_K \,.
    \label{eq:nR1upp1}
\end{align}
Furthermore, since $K$ is independent of $(M_0,M_1)$, 
\begin{align}
    I(M_1;B^n,K|M_0)_\rho=I(M_1;B^n|M_0,K)_\rho \\
    I(M_1;E^n,K|M_0)_\rho=I(M_1;E^n|M_0,K)_\rho \,.
    \label{eq:nR1upp2}
\end{align}
By inserting (\ref{eq:nR1upp1})-(\ref{eq:nR1upp1}) into
(\ref{eq:nR1upp0k}), we have
\begin{align}
    nR_1&\leq  I(M_1;B^n|M_0,K)_\rho-I(M_1;E^n|M_0,K)_\rho+nR_K+n\delta_n+n\eps_{3n} \,.
    \label{eq:nR1kU3}
\end{align}

Defining $T^n=f_0(M_0,K)$ and  $X^n=f_1(M_0,M_1,K)$, where $f_j$ are one-to-one mappings, we obtain
\begin{align}
    R_0&\leq \frac{1}{n}\min\left( I(T^n;B^n )_{\rho}+\eps_{2n} \,,\; I(T^n;E^n )_{\rho}+\eps_{1n}  \right)\\
    R_1&\leq \frac{1}{n}\min\left( [I(X^n;B^n|T^n)_\rho-I(X^n;E^n|T^n)_\rho]_{+}+R_K+\delta_n +\eps_{3,n} \,,\; I(X^n;B^n|T^n)_\rho+\eps_{3,n} \right)
\end{align}
based on (\ref{eq:R0convEk}), (\ref{eq:R0convBk}), (\ref{eq:nR1kU2}), and (\ref{eq:nR1kU3}). This completes the regularized converse proof.
\qed

\section{Proof of Theorem~\ref{theo:distillationF}}
\label{app:distillationF}
Consider the key-agreement protocol for the distillation of a public key $k_0$ and a secret key $k_1$ from a given quantum state $\omega_{ABE}^{\otimes n}$.  
\subsection{Achievability proof}
We modify and extend Devetak and Winter's methods. 
When Alice performs the measurement $\Lambda$ on each of her systems $A_i$, she obtains a classical memoryless source sequence $X^n$, which is i.i.d. $\sim p_X(x)=\trace(\Lambda_A^x \omega_A)$.
Roughly speaking,  one may encode this source using indices, such that this source sequence ``appears" to Bob and Eve as a codeword for the  classical-quantum-quantum broadcast channel $\channel_{X\rightarrow BE}$ that corresponds to $\omega_{XBE}$.
Our key-agreement coding schemes is specified below in further details.

Fix the POVM $\Lambda_A$ and the conditional distribution $p_{T_0,T_1|X}$. Denote the joint distribution of $X$, $T_0$, and $T_1$ by
  \begin{align}
      p_{X,T_0,T_1}(x,t_0,t_1)=
      \trace(\Lambda_A^x\omega_A)p_{T_0,T_1|X}(t_0,t_1|x) \,.
  \end{align}
  A key-agreement code is constructed as follows.

\subsubsection*{Classical codebook construction}
For every given joint type $\hP_{X,T_0,T_1}$ on $\Xset\times\Tset_0\times\Tset_1$, select 
$2^{n(\tR_0+R_0)}$ independent sequences $U^{\ell_0,k_0}$, $\ell_0\in [1:2^{n\tR_0}]$, $k_0\in [1:2^{nR_0}]$, at random, each is uniformly distributed over the type class $\mathscr{T}(n,\hP_{T_0})$.
Furthermore, select 
$2^{n(\tR_1+R_1+R_s)}$ independent sequences $V^{\ell_1,k_1,s}$, $\ell_1\in [1:2^{n\tR_1}]$, $k_1\in [1:2^{nR_1}]$, and $s\in [1:2^{nR_s}]$, at random, each is uniformly distributed over the conditional type class $\mathscr{T}(n,\hP_{X|T_0,T_1})$. 

\subsubsection*{Encoding}
Alice measures the system $A^n$ using the POVM $\Lambda_A$. 
  Given the measurement outcome $x^n$, she generates the random sequences
  $(t_0^n,t_1^n)\sim \prod_{i=1}^n p_{T_0,T_1|X}(t_{0,i},t_{1,i}|x_i)$. The resulting state is $\omega_{X T_0 T_1 BE}^{\otimes n}$. 
  
  Alice computes the joint type $\hP\equiv \hP_{x^n,t_0^n,t_1^n}$.
  If the tuple $(x^n,t_0^n,t_1^n)$ is not $\delta$-typical, \ie 
  $\min_{x,t_0,t_1}|\hP(x,t_0,t_1
  )-p_{X,T_0,T_1}(x,t_0,t_1)|>\delta$, then the protocol aborts.
 Otherwise, she sends the type $\hP$ to Bob.
Then, Alice chooses 
$\ell_1,k_1,s$ at random such that $V^{\ell_1,k_1,s}=x^n$, and informs Bob of $\ell_1$ and  $t_0^n$, $t_1^n$ as well. 
  Similarly, Alice chooses 
$\ell_0,k_0$ at random such that $U^{\ell_0,k_0}=t_0^n$, and informs Eve of $\ell_0$ and the type of $t_0^n$. 

\subsubsection*{Decoding and key generation}
Alice sets her key as $(K_0,K_1)=(k_0,k_1)$.
Eve and Bob receive  $\ell_0$ and   $(\ell_1,t_0^n,t_1^n)$, respectively, along with the respective types. They perform  measurements using the respective  POVMs $\{ \Xi_{E^n|\ell_0,\hP}^{k_0} \}$ and
$\{\Gamma_{B^n|\ell_1,\hP}^{k_1,s}\}_{k_1,s}$,
which will be specified later.
Eve and Bob obtain the measurement outcomes, $\tk_0$ and $\hk_0,\hk_1,\hs$, and set their keys as $\tK_0=\tk_0$ and
$(\hK_0,\hK_1)=(\hk_0,\hk_1)$, respectively.

\subsubsection*{Error analysis}
 Denote 
\begin{align}
    \sigma^{t_0^n}_{B^n E^n}(\hP)=\frac{1}{|\Tset(\hP)|} \sum_{(x^n,t_1^n)\in\Tset(\hP)} \omega_{B^n E^n}^{x^n,t_0^n,t_1^n} 
\end{align}
where $\omega_{B^n E^n}^{x^n,t_0^n,t_1^n}=\bigotimes_{i=1}^n \omega_{B E}^{x_i,t_{0,i},t_{1,i}} $ with 
\begin{align}
    \omega_{B E}^{x,t_0,t_1}\equiv 
    (\bra{x}\otimes\bra{t_0}\otimes\bra{t_1}\otimes\identity)\omega_{XT_0T_1BE}(\ket{x}\otimes\ket{t_0}\otimes\ket{t_1}\otimes\identity)
    \,.
\end{align}
We define the averaged state $\omega^{t_0}_{BE}=
\sum_{x,t_1} p_{X,T_1|T_0}(x,t_1|t_0)\omega_{BE}^{x,t_0,t_1}$, hence  $\omega_{B^n E^n}^{t_0^n}=\bigotimes_{i=1}^n \omega_{B E}^{t_{0,i}}$; and in a similar manner, we also define
$\omega^{x}_{BE}$ and $\omega_{B^n E^n}^{x^n}$.

Next, we use the classical capacity theorem for a classical-quantum channel.
According to the modified HSW Theorem \cite[Proposition 5]{DevetakWinter:05p},  for every given $\ell_0$, there exists a POVM $\{\Xi_{E^n|\ell_0,\hP}^{k_0}\}_{k_0}$ that guarantees reliable decoding, \ie such that 
\begin{align}
    \trace\left( \Xi_{E^n|\ell_0,\hP}^{k_0}\, \omega_{E^n}^{U^{\ell_0,k_0}} \right)\geq 1-\eps_0
\end{align}
 when $n$ is sufficiently large, provided that 
\begin{align}
    R_0&< I(T_0;E)_\omega-\eps_1
    \label{eq:R1u1DirD}
\end{align}
 where $\eps_j>0$ are arbitrarily small.
 Thus, Eve can recover $k_0$ reliably using this POVM.

In the same manner, %
there exists a POVM $\{\Gamma_{B^n|\ell_1,\hP}^{k_0,k_1,s_1}\}_{k_0,k_1,s_1}$ such that 
\begin{align}
    \trace\left( \Gamma_{B^n|\ell_1,\hP}^{k_0,k_1,s_1}\, \omega_{B^n}^{V^{\ell_1,k_0,k_1,s_1}} \right)\geq 1-\eps_0
\end{align}
 when $n$ is sufficiently large, provided that 
\begin{align}
    R_1+R_s&< I(X,T_0,T_1;B,T_0,T_1)_\omega-\eps_1
    \nonumber\\
    &=I(X;B|T_0,T_1)_\omega-\eps_1
    \label{eq:R1u1DirDa}
\end{align}
 where $\eps_j>0$ are arbitrarily small.
 Thus, Bob can recover $k_0,k_1,s$ reliably using this POVM.

\subsubsection*{Secrecy and rate analysis}
As for the secrecy, according to the covering lemma 
\cite[Proposition 4]{DevetakWinter:05p}, 
\begin{align}
    \Pr\left( \norm{ \frac{1}{2^{nR_S}} \sum_{s=1}^{2^{nR_S}}\omega_{E^n}^{V^{\ell,k_1,s}}-\sigma^{t_0^n}_{E^n}(\hP) }_1>\delta_1 \right)\leq 
    2^{-2^{n(R_S-I(X;E|T_1,T_0)_\omega-\eps_2))}}
\end{align}
where $\delta_1,\eps_2>0$ are arbitrarily small. Thus, Eve's state is $\delta_1$-close to a constant state that does not depend on $\ell_1,k_1$ with double-exponentially high probability, provided that 
\begin{align}
    R_s>I(X;E|T_0,T_1)_\omega+\eps_2 \,.
    \label{eq:R1u2DirD}
\end{align}

We have shown that the key $k_0$ can be distributed between Alice, Bob, and Eve, provided that $R_0<I(T_0;E)_\rho-\eps_1$
(see (\ref{eq:R1u1DirD})). 
Furthermore, based on (\ref{eq:R1u1DirD}) and (\ref{eq:R1u2DirD}), the key $k_1$ can be distributed confidentially provided that $R_1<I(X;B|T_0,T_1)_\omega-I(X;E|T_0,T_1)_\omega$. 
Since each codebook is restricted to a particular type class, $k_0$ and $k_1$ are uniformly distributed, hence 
$H(K_j)\geq nR_j$ for $j=0,1$.
This completes the achievability proof.

\subsection{Converse proof}
Suppose that Alice, Bob, and Eve share a quantum state $\omega_{ABE}^{\otimes n}$. Alice distills the public and confidential keys $(K_0,K_1)$ by measuring her system, $A^n$. 
 She sends a classical message $z_b$ to Bob, and a classical message $z_e$ to Eve  through a public channel.  The output state is
$\rho_{K_0 K_1 Z_b Z_e B^n E^n}$.
Then, Bob and Eve use the messages that they have received and perform POVMs $\Gamma^{k_0,k_1}_{B^n|z_b}$ and $\Xi^{k_0}_{E^n|z_e}$, respectively. Doing so, Bob obtains a pair of public and confidential keys, $(\hK_0,\hK_1)$, as measurement outcomes, and Eve obtains the public key $\tK_0$.
Consider a sequence of codes $(F_n,\Lambda_n,\Gamma_n)$ such that the key rates satisfy
\begin{align}
    \frac{1}{n} H(K_j)\geq n(R_j-\alpha_n) \,,\; j=0,1;
    \label{eq:convRjkey01}
\end{align}
and the average probability of error and leakage rates tend to zero. 
By Fano's inequality \cite{CoverThomas:06b}, it follows that%
\begin{align}
H(K_0|\tK_0) \leq n\eps_n
\label{eq:Hm0ud}
\\
H(K_0|\hK_0) \leq n\eps_n'
\label{eq:Hm0udtk}
\\
H(K_1|\hK_1) \leq n\eps_n''
\label{eq:Hm1udtk}
\end{align}
where $\eps_n,\eps_n',\eps_n''$ tend to zero as $n\rightarrow\infty$.
Since the leakage rates tend to zero, we also have
\begin{align}
I(Z_b Z_e;K_0)_\rho\leq n\delta_n
\label{eq:LconverseD0}
\\
I(Z_b Z_e E^n;K_1)_\rho\leq n\delta_n
\label{eq:LconverseD1}
\end{align}
where $\delta_n$ tends to zero as $n\rightarrow\infty$.

Thus, we bound the public key rate by
\begin{align}
    n(R_0-\alpha_n)&\leq H(K_0)
    \nonumber\\
    &\leq H(K_0)-H(K_0|\tK_0) + n\eps_n
    \nonumber\\
    &= I(K_0;\tK_0)+ n\eps_n
    \nonumber\\
    &\leq I(K_0;Z_e,E^n)_\rho+ n\eps_n
    \label{eq:K0Convu1}
\end{align}
where the first inequality holds by (\ref{eq:convRjkey01}), the second inequality follows from (\ref{eq:Hm0ud}), and the last inequality is due to the data processing inequality for the quantum mutual information. 
Based on the leakage requirement for the public key, 
$I(Z_e;K_0)_\rho\leq I(Z_b,Z_e;K_0)_\rho\leq \delta_n $ (see (\ref{eq:LconverseD0})). Thus,
\begin{align}
    I(K_0;Z_e,E^n)_\rho&\leq 
    I(K_0;E^n|Z_e)_\rho+n\delta_n
    \nonumber\\
    &\leq I(K_0,Z_e,Z_b;E^n)_\rho+n\delta_n
\end{align}
Together with (\ref{eq:K0Convu1}), this implies
\begin{align}
    R_0\leq \frac{1}{n} I(K_0,Z_e,Z_b;E^n)_\rho+\alpha_n+\eps_n+\delta_n \,.
    \label{eq:Dr0E}
\end{align}
By applying the same arguments to Bob, we have
\begin{align}
    R_0\leq \frac{1}{n} I(K_0,Z_e,Z_b;B^n)_\rho+\alpha_n+\eps_n'+\delta_n \,.
    \label{eq:Dr0B}
\end{align}

We continue to the confidential key.
Notice that since the keys are classical, we have
\begin{align}
    I(K_1;K_0|Z_e,Z_b,E^n)_\rho
    &\leq H(K_0|Z_e,Z_b,E^n)_\rho
    \nonumber\\
    &= H(K_0)-I(K_0;Z_e,Z_b,E^n)_\rho\leq n\eps_n
\end{align}
where the last inequality follows as in (\ref{eq:K0Convu1}). Adding the last bound to (\ref{eq:LconverseD1}), this yields
\begin{align}
    I(K_1;K_0,Z_e,Z_b,E^n)_\rho\leq n(\eps_n+\delta_n) \,.
    \label{eq:Ik0k1u}
\end{align}
Then, the confidential key rate is bounded as
\begin{align}
    n(R_1-\alpha_n)
    &\leq I(K_1;Z_b,B^n)_\rho+ n\eps_n''
    \nonumber\\
    &\leq I(K_1;K_0,Z_e,Z_b,B^n)_\rho+ n\eps_n''
     \nonumber\\
     &\leq I(K_1;K_0,Z_e,Z_b,B^n)_\rho
     -I(K_1;K_0,Z_e,Z_b,E^n)_\rho
     + n(\eps_n+\delta_n+\eps_n'')
         \nonumber\\
     &= I(K_1;B^n|K_0,Z_e,Z_b)_\rho
     -I(K_1;E^n|K_0,Z_e,Z_b)_\rho
     + n(\eps_n+\delta_n+\eps_n'')
    \label{eq:K1Convu1}
\end{align}
where the first inequality is based on similar arguments as we used in order to show (\ref{eq:K0Convu1}), the third inequality holds by (\ref{eq:Ik0k1u}), and the last equality comes from the chain rule.
The proof for the regularized converse part follows from (\ref{eq:Dr0E})-(\ref{eq:Dr0B}) and (\ref{eq:K1Convu1}),  
by defining  $T_0^n=T_1^n=f_0(K_0,Z_e,Z_b)$ and  $X^n=f_1(K_1)$, where $f_j$ are one-to-one mappings. 
This completes the proof of Theorem~\ref{theo:distillationF}.
\qed

\section{Proof of Theorem~\ref{theo:LS}}
\label{app:LS}
Consider layered-secrecy communication over the degraded broadcast channel $\channel_{ABE_1 E_2}$.

\subsection{Achievability proof}
We show that for every $\zeta_0,\eps_0,\delta_0>0$, there exists a $(2^{n(R_0-\zeta_0)},2^{n(R_1-\zeta_0)},2^{n(R_2-\zeta_0)},n,\eps_0,\delta_0)$ layered-secrecy code for the quantum broadcast channel  $\channel_{A\rightarrow B E_1 E_2}$, provided that $(R_0,R_1,R_2)\in \mathcal{R}_{\text{LS}}(\channel)$. 
To prove achievability, we extend  the classical combination of super-position coding with random binning, and then apply the quantum packing lemma and the quantum covering lemma. We use the gentle measurement lemma \cite{Winter:99p}, %
which guarantees that multiple decoding measurements can be performed without ``destroying" the output state.

\subsection*{Useful lemmas}
We make heavy use of the quantum packing lemma, quantum covering lemma, and gentle-measurement lemma. 
Those lemmas are given below.

We begin with the quantum packing lemma, which is a useful tool in proofs of  channel coding theorems.
\begin{lemma}[Quantum Packing Lemma {%
\cite{HsiehDevetakWinter:08p}\cite[Corollary 16.5.1]{Wilde:17b}}]
\label{lemm:Qpacking}
Let %
\begin{align}
\rho=\sum_{x\in\Xset} p_X(x) \rho_x
\end{align}
where $\{ p_X(x), \rho_x \}_{x\in\Xset}$ is a given ensemble. %
Furthermore, suppose that there is  a code projector $\Pi$ and codeword projectors $\Pi_{x^n}$, $x^n\in\tset(p_X)$, that satisfy for every 
$\alpha>0$ and sufficiently large $n$,
\begin{align}
\trace(\Pi\sigma_{x^n})\geq&\, 1-\alpha \\
\trace(\Pi_{x^n}\sigma_{x^n})\geq&\, 1-\alpha \\
\trace(\Pi_{x^n})\leq&\, 2^{n e_0}\\
\Pi \rho^{\otimes n} \Pi \preceq&\, 2^{-n(E_0-\alpha)} \Pi 
\end{align}
for some $0<e_0<E_0$ with $\sigma_{x^n}\equiv \bigotimes_{i=1}^n \rho_{x_i}$.
Consider a classical random codebook $\mathscr{C}=\{ X^n(m)$, $m\in [1:2^{nR}] \}$, that consists 
of independent sequences, each i.i.d. $\sim p_X$.
Then, there exists   a POVM $\{ \Lambda_m \}_{m\in [1:2^{nR}]}$ such that 
\begin{align}
\label{eq:QpackB}
  \mathbb{E}_{\mathscr{C}}[\trace\left( \Lambda_m \sigma_{X^n(m)} \right)]  \geq 1-2^{-n[ E_0-e_0-R-\eps_n(\alpha)]}
\end{align}
for all %
$m\in [1:2^{nR}]$, where $\eps_n(\alpha)$ tends to zero as $n\rightarrow\infty$ and $\alpha\rightarrow 0$. 
\end{lemma}

Next, we give the quantum covering lemma, which originated from   source coding analysis.
\begin{lemma}[Quantum Covering Lemma {%
\cite[Lemma 17.2.1]{Wilde:17b}}]
\label{lemm:Qcovering}
Fix $\delta>0$. Let 
\begin{align}
\rho=\sum_{x\in\Xset} p_X(x) \rho_x
\end{align}
where $\{ p_X(x), \rho_x \}_{x\in\Xset}$ is a given ensemble. %
Furthermore, suppose that there is  a code projector $\Pi$ and codeword projectors $\Pi_{x^n}$, $x^n\in\tset(p_X)$, that satisfy for every 
$\alpha>0$ and sufficiently large $n$,
\begin{align}
\trace(\Pi\sigma_{x^n})\geq&\, 1-\alpha \\
\trace(\Pi_{x^n}\sigma_{x^n})\geq&\, 1-\alpha \\
\trace(\Pi)\leq&\, 2^{n E_0}\\
\Pi_{x^n} \sigma_{x^n} \Pi_{x^n} \preceq&\, 2^{-n(e_0-\alpha)} \Pi 
\end{align}
for some $0<e_0<E_0$ with $\sigma_{x^n}\equiv \bigotimes_{i=1}^n \rho_{x_i}$.
Consider a classical random codebook $\mathscr{C}=\{ X^n(m)$, $m\in [1:2^{nR}] \}$, that consists 
of independent sequences, each i.i.d. $\sim p_X$. 
Then, 
\begin{align}
\label{eq:QcoverB}
  \prob{\norm{\rho^{\otimes n}-\frac{1}{2^{nR}}\sum_{m=1}^{2^{nR}}\sigma_{X^n(m)}}_1>\delta}  \leq \exp\left(-2^{n[R- E_0+e_0-\eps_n(\alpha)]}\right)
\end{align}
 where $\eps_n(\alpha)$ tends to zero as $n\rightarrow\infty$ and $\alpha\rightarrow 0$. 
\end{lemma}

As will be seen, the gentle measurement lemma guarantees that we can perform multiple measurements such that the 
state of the system remains almost the same after each measurement.

\begin{lemma}[see {\cite{Winter:99p,OgawaNagaoka:07p}}]
\label{lemm:gentleM}
Let $\rho$ be a density operator. Suppose that $\Lambda$ is a meaurement operator such that $0\preceq \Lambda\preceq \identity$. If
\begin{align}
\trace(\Lambda\rho) \geq 1-\eps
\end{align}
for some $0\leq\eps\leq 1$, then the post-measurement state $\rho'\equiv \frac{\sqrt{\Lambda}\rho\sqrt{\Lambda} }{\trace(\Lambda\rho)}$ is $2\sqrt{\eps}$-close to the original state in trace distance, \ie
\begin{align}
\norm{ \rho-\rho' }_1\leq 2\sqrt{\eps} \,.
\end{align}
\end{lemma}
The lemma is particularly useful in our analysis since the POVM operators in the quantum packing lemma satisfy the conditions of the lemma for large $n$ (see (\ref{eq:QpackB})).

\subsection*{Quantum Method of Types}
Standard method-of-types concepts are defined as in  \cite{Wilde:17b,Pereg:19a3}.  
We briefly introduce the notation and basic properties while the detailed definitions can be found in \cite[Appendix A]{Pereg:19a3}.
In particular, given a density operator $\rho=\sum_x p_X(x)\ketbra{x}$ on the Hilbert space $\Hset_A$, we let
$\tset(p_X)$ denote the $\delta$-typical set that is associated with $p_X$, and
 $\Pi_{A^n}^{\delta}(\rho)$ the projector onto the corresponding subspace.  
The following inequalities follow from well-known properties of $\delta$-typical sets \cite{NielsenChuang:02b}, %
\begin{align}
\trace( \Pi^\delta(\rho) \rho^{\otimes n} )\geq& 1-\eps  \label{eq:UnitT} \\
 2^{-n(H(\rho)+c\delta)} \Pi^\delta(\rho) \preceq& \,\Pi^\delta(\rho) \,\rho^{\otimes n}\, \Pi^\delta(\rho) \,
\preceq 2^{-n(H(\rho)-c\delta)}
\label{eq:rhonProjIneq}
\\
\trace( \Pi^\delta(\rho))\leq& 2^{n(H(\rho)+c\delta)} \label{eq:Pidim}
\end{align}
 where $c>0$ is a constant.
Furthermore, for $\sigma_B=\sum_x p_X(x)\rho_B^x$, %
let $\Pi_{B^n}^{\delta}(\sigma_B|x^n)$ denote the projector corresponding to the conditional $\delta$-typical set %
given the sequence $x^n$.
Similarly \cite{Wilde:17b}, %
\begin{align}
\trace( \Pi^\delta(\sigma_B|x^n) \rho_{B^n}^{x^ n} )\geq& 1-\eps'  \label{eq:UnitTCond} \\
 2^{-n(H(B|X')_\sigma+c'\delta)} \Pi^\delta(\sigma_B|x^n) \preceq& \,\Pi^\delta(\sigma_B|x^n) \,\rho_{B^n}^{x^ n}\, \Pi^\delta(\sigma_B|x^n) \,
\preceq 2^{-n(H(B|X')_{\sigma}-c'\delta)}
\label{eq:rhonProjIneqCond}
\\
\trace( \Pi^\delta(\sigma_B|x^n))\leq& 2^{n(H(B|X')_\sigma+c'\delta)} \label{eq:PidimCond}
\end{align}
where $c'>0$ is a constant, $\rho_{B^n}^{x^n}=\bigotimes_{i=1}^n \rho_{B_i}^{x_i}$, and the classical random variable $X'$ is distributed according to the type of $x^n$.
If $x^n\in\tset(p_X)$, then %
\begin{align}
\trace( \Pi^\delta(\sigma_B) \rho_{B^n}^{x^n} )\geq& 1-\eps' \,. 
\label{eq:UnitTCondB}
\end{align}
 as well (see \cite[Property 15.2.7]{Wilde:17b}).
We note that the conditional entropy in the bounds above can also be expressed as 
$%
H(B|X')_\sigma=\frac{1}{n} H(B^n|X^n=x^n)_{\sigma}  \equiv \frac{1}{n} H(B^n)_{\rho^{x^n}} %
$. %

\subsection*{Coding Scheme}
The code construction, encoding and decoding procedures are described below.
Let $\{ p_{X_0,X_1,X_2} , \varphi_{A}^{x_0,x_1,x_2} \}$ be a given ensemble. %
Consider 
\begin{align}
    \rho_{BE_1 E_2}^{x_0,x_1,x_2}\equiv \channel_{A\rightarrow BE_1 E_2}(\varphi_{A}^{x_0,x_1,x_2}) \,.
\end{align}
It will be useful for use to define the averaged state,
 \begin{align}
     \rho_{B E_1 E_2}^{x_0,x_1}=\sum_{x_2\in\Xset_2} p_{X_2|X_1,X_0}(x_2|x_1,x_0) \rho_{BE_1 E_2}^{x_0,x_1,x_2} %
 \end{align}
 having averaged over $x_2$, for a given $x_0$ and $x_1$.
The layered-secrecy code construction is defined as follows.

\vspace{0.25cm}
\subsubsection{Classical Code Construction}
Let $\tR_j>R_j$ for $j=1,2$, and $\delta>0$.  
Select $2^{nR_0}$ independent sequences $x_0^n(m_0)$, $m_0\in [1:2^{nR_0}]$, at random, each according to $\prod_{i=1}^n p_{X_0}(x_{0,i})$.
Then, for every given $x_0^n(m_0)$, do as follows. Generate $2^{nR_1}$ subcodebooks $\mathscr{C}_1(m_0,m_1)$, $m_1\in[1:2^{nR_1}]$, each consists of 
$2^{n(\tR_1-R_1)}$ conditionally independent random sequences,
\begin{align}
   x_1^n(m_0,\ell_1) \,,\; \ell_1\in [(m_1-1)2^{n(\tR_1-R_1)}+1:m_1 2^{n(\tR_1-R_1)}] 
\end{align}
drawn according to $ \prod_{i=1}^n p_{X_1|X_0}(x_{1,i}|x_{0,i}(m_0))$.
Next, for every $\ell_1$, 
generate a subcodebook $\mathscr{C}_2(m_0,\ell_1,m_2)$ that consists of 
$2^{n(\tR_2-R_2)}$ conditionally independent random sequences,
\begin{align}
    x_2^n(m_0,\ell_1,\ell_2) \,,\;\ell_2\in [(m_2-1)2^{n(\tR_2-R_2)}+1:m_2 2^{n(\tR_2-R_2)}] 
\end{align}
 drawn according to $ \prod_{i=1}^n p_{X_2|X_1,X_0}(x_{2,i}|x_{1,i}(m_0,\ell_1),x_{0,i}(m_0))$.

\vspace{0.25cm}
\subsubsection{Encoding}
To send the message tuple $(m_0,m_1,m_2)$, Alice performs the following.
\begin{enumerate}[(i)]
\item
Select $\ell_j$  uniformly at random from $ [(m_j-1)2^{n(\tR_j-R_j)}+1:m_j 2^{n(\tR_j-R_j)}]$, for $j=1,2$.

\item
Prepare 
\begin{align}
\sigma_{A^n}^{m_0,\ell_1,\ell_2}= \bigotimes_{i=1}^n \rho_A^{x_{0,i}(m_0),x_{1,i}(m_0,\ell_1),x_{2,i}(m_0,\ell_1,\ell_2))  }    
\end{align}
 and send the input system $A^n$.
\end{enumerate}

\vspace{0.25cm}
\subsubsection{Decoding}
Bob, Eve 1, and Eve 2 receive the output systems $B^n$, $E_1^n$, and $E_2^n$ in the state 
\begin{align}
\sigma^{m_0,\ell_1,\ell_2}_{B^{n} E_1^n E_2^n}= \bigotimes_{i=1}^n \rho_{BE_1 E_2}^{x_{0,i}(m_0),x_{1,i}(m_0,\ell_1),x_{2,i}(m_0,\ell_1,\ell_2)  }
\label{eq:rhoBTnSC}
\end{align}
 and decode as follows. 

Eve 2 decodes $\breve{m}_0$ by applying a POVM $\{ \Upsilon_{m_0} \}_{
m_0 \in  [1:2^{R_0}]}$, which will be specified later, to the system $E_2^n$. 

Eve 1 also decodes $\tm_0$ by applying a POVM $\{ \Xi'_{m_0} \}_{
m_0 \in  [1:2^{R_0}]}$. Then, she decodes $\tilde{\ell}_1$ by applying a second POVM $\{ \Xi''_{\ell_1|\tm_0} \}_{
\ell_1 \in  [1:2^{\tR_1}]}$, which will also be specified later, to the system $E_1^n$. She declares that the message  $\tm_1$ was sent, where $\tm_1$ is the  subcodebook index that is associated with $\tilde{\ell}_1$, \ie 
\begin{align}
x_1^n(\tm_0,\tilde{\ell}_1)\in \mathscr{C}_1(\tm_0,\tm_1) \,.    
\end{align}

Similarly, Bob decodes $\hm_0$, $\hat{\ell}_1$, and $\hat{\ell}_2$, by applying three consecutive POVMs,
$\{ \Gamma_{m_0}' \}_{m_0 \in  [1:2^{R_0}] }$, 
$\{ \Gamma_{\ell_1|\hm_0}'' \}_{\ell_1 \in  [1:2^{\tR_1}] }$, and %
$\{ \Gamma_{\ell_2|\hm_0,\hat{\ell}_1}'' \}_{\ell_2 \in  [1:2^{\tR_2}] }$, which will also be specified later.
He declares  $(\hm_0,\hm_1,\hm_2)$ as the subcodebook indices such that 
\begin{align}
    x_1^n(\hm_0,\hat{\ell}_1)\in \mathscr{C}_1(\hm_0,\hm_1)
    \,\text{ and }\;
x_2^n(\hm_0,\hat{\ell}_1,\hat{\ell}_2)\in \mathscr{C}_2(\hm_0,\hat{\ell}_1,\hm_2)
\end{align}
 hold simultaneously.

\vspace{0.25cm}
\subsubsection{Analysis of Probability of Error and Layered Secrecy}
By symmetry, we may assume without loss of generality that Alice sends the messages $M_0=M_1=M_2=1$ using $L_1=L_2=1$. 

Consider the following probabilistic events,
\begin{align}
\mathscr{G}_A&= \{  (X_0^n(1),X_1^n(1,1),X_2^n(1,1,1))\notin \Aset^{\delta_1}(p_{X_0,X_1,X_2})  \} 
\intertext{and}
\mathscr{D}_{B,0}&= \{  \hM_0 \neq 1  \}\\
\mathscr{D}_{E_1,0}&= \{  \widetilde{M}_0 \neq 1  \}\\
\mathscr{D}_{E_2,0}&= \{  \breve{M}_0 \neq 1  \}\\
\mathscr{D}_{B,1}&= \{  \hL_1 \neq 1  \}\\
\mathscr{D}_{E_1,1}&= \{  \widetilde{L}_1 \neq L_1  \}\\
\mathscr{D}_{B,2}&= \{  \hL_2 \neq 1  \}
\end{align}
with $\delta_1\equiv \delta/(2|\Xset_1| |\Xset_2| )$, where the notation 
$\mathscr{D}_{s,j}$ indicates the decoding error of Receiver $s$ with respect to the layer-$j$ message.
We also consider the secrecy violation events,
\begin{align}
    \mathcal{S}_{E_2,1}=& \{  \norm{ \sigma_{E_2^n}^{\ell_1}-\breve{\sigma}_{E_2^n} }_1>\delta  \}
\\
\mathcal{S}_{E_1 E_2,2}=& \{  \norm{ \sigma_{E_1^n E_2^n}^{\ell_1,\ell_2}-\widetilde{\sigma}^{\ell_1}_{E_1^n E_2^n} }_1>\delta  \}
\end{align}
where we have denoted the averaged output states by
\begin{align}
    \breve{\sigma}_{B^n E_1^n E_2^n}&=
    \frac{1}{2^{n(\tR_1+\tR_2)}}\sum_{\ell_1=1}^{2^{n\tR_1}}\sum_{\ell_2=1}^{2^{n\tR_2}}\sigma^{\ell_1,\ell_2}_{B^n E_1^n E_2^n}
    \intertext{and}
    \widetilde{\sigma}^{\ell_1}_{B^n E_1^n E_2^n}&=
    \frac{1}{2^{n\tR_2}}\sum_{\ell_2=1}^{2^{n\tR_2}}\sigma^{\ell_1,\ell_2}_{B^n E_1^n E_2^n} \,.
\end{align}

 By the union of events bound, the probability of error is bounded by
 \begin{align}
 P_{e|m_1=1,m_2=1}^{(n)}(\Fset,\Gamma,\Xi,\Upsilon) &\leq \prob{ \mathscr{G}_A }+
 \prob{ \mathscr{D}_{E_2,0} |\mathscr{G}_A^c}+
 \prob{ \mathscr{D}_{E_1,0} |\mathscr{G}_A^c}+
 \prob{ \mathscr{D}_{B,0} |\mathscr{G}_A^c} \nonumber\\
 &+
 \prob{ \mathscr{D}_{E_1,1} |\mathscr{G}_A^c\cap \mathscr{D}_{E_1,0}^c}+\prob{ \mathcal{S}_{E_2,1} |\mathscr{G}_A^c\cap \mathscr{D}_{E_2,0}^c
 }
 \nonumber\\ &+
 \prob{ \mathscr{D}_{B,2} |\mathscr{G}_A^c\cap \mathscr{D}_{B,0}^c\cap \mathscr{D}_{B,1}^c}+
 \prob{ \mathcal{S}_{E_1 E_2,2} |\mathscr{D}_{E_1,0}^c\cap
 \mathscr{D}_{E_2,0}^c\cap \mathscr{D}_{E_1,1}^c}
 \label{eq:PeBsc}
 \end{align}
 where the conditioning on $M_j=1$ and $L_j=1$ is omitted for convenience of notation.
The first term tends to zero as $n\rightarrow\infty$ by the law of large numbers. 

Eve 2's error event for the common message $M_0$ corresponds to
the second term on the RHS of (\ref{eq:PeBsc}).
To bound this term, we use the quantum packing lemma.
  Given that the event $\mathscr{A}^c$ has occurred, we have $X_0^n(1)\in\Aset^{
 \nicefrac{\delta}{2}}(p_{X_0})$. %
 Now, by the basic properties of type class projectors, %
 \begin{align}
 \Pi^{\delta}(\rho_{E_2})  \rho_{E_2^n}   \Pi^{\delta}(\rho_{E_2}) \preceq& 2^{ -n(H(E_2)_{\rho}-\eps_1(\delta)) } \Pi^{\delta}(\rho_{E_2})
 \\
 \trace\left[ \Pi^{\delta}(\rho_{E_2}|x_0^n) \rho_{E_2^n}^{x_0^n} \right] \geq& 1-\eps_1(\delta) \\
 \trace\left[ \Pi^{\delta}(\rho_{E_2}|x_0^n)  \right] \leq& 2^{ n(H(E_2|X_0)_{\rho} +\eps_1(\delta))} \\
 \trace\left[ \Pi^{\delta}(\rho_{E_2}) \rho_{E_2^n}^{x_0^n} \right] \geq& 1-\eps_1(\delta) 
 \end{align}
 for all $x_0^n\in\Aset^{\delta_1}(p_{X_0})$, by (\ref{eq:rhonProjIneq}), (\ref{eq:UnitTCond}), (\ref{eq:PidimCond}), and (\ref{eq:UnitTCondB}), respectively, where $\eps_i(\delta)\rightarrow 0$ as $\delta\rightarrow 0$.
  By the quantum packing lemma, Lemma~\ref{lemm:Qpacking}, there exists a POVM $\Upsilon_{m_0}$, for Eve 2, such that
 \begin{align}
 \cprob{ \mathscr{D}_{E_2,0} }{ \mathscr{G}_A^c } \leq 2^{ -n( I(X_0;E_2)_\rho -R_0-\eps_2(\delta)) } \,.
 \end{align}
 The last expression tends to zero as $n\rightarrow\infty$, provided that 
 \begin{align}
 R_0< I(X_0;E_2)_\rho -\eps_2(\delta) \,.
 \label{eq:E2}
 \end{align}

Consider the layer-0 error events for Eve 1 and Bob, $\mathscr{D}_{E_1,0}$ and $\mathscr{D}_{B,0}$, respectively.
Applying the same argument to the output systems $E_1^n$ and $B^n$, we find that there exist respective POVMs $\Xi_{m_0}'$ and $\Gamma_{m_0}'$, for Eve 1 and Bob,  such that
 \begin{align}
 \cprob{ \mathscr{D}_{E_1,0} }{ \mathscr{G}_A^c } &\leq 2^{ -n( I(X_0;E_1)_\rho -R_0-\eps_2(\delta)) }
 \label{eq:error0boundE1}
 \\
 \cprob{ \mathscr{D}_{B,0} }{ \mathscr{G}_A^c } &\leq 2^{ -n( I(X_0;B)_\rho -R_0-\eps_2(\delta)) }
 \label{eq:error0boundB}
 \end{align}
 for sufficiently large $n$.
 We claim that the last expression vanishes if (\ref{eq:E2}) holds.
Indeed,  since the channel is degraded, %
 \begin{align}
  I(X_0;E_2)_\rho\leq I(X_0;E_1)_\rho\leq I(X_0;B)_\rho 
 \label{eq:BdegD1}
 \end{align}
by the data processing inequality. Thus, (\ref{eq:E2}) implies 
$R_0< I(X_0;E_1)_\rho -\eps_2(\delta)\leq I(X_0;B)_\rho -\eps_2(\delta)$. This, in turn, implies that the error probability $ \cprob{ \mathscr{D}_{E_1,0} }{ \mathscr{A}^c }$ and $ \cprob{ \mathscr{D}_{B,0} }{ \mathscr{A}^c }$ tend to zero as $n\rightarrow\infty$ by (\ref{eq:error0boundE1})-(\ref{eq:error0boundB}).

We move to the decoding errors for the layer-1 message. Let $\sigma'^{\, m_0,\ell_1,\ell_2 }_{E_1^n}$ denote the state of Eve 1's output system after applying the measurement $\Xi'_{m_0}$ above. Based on the  gentle measurement lemma, Lemma~\ref{lemm:gentleM}, and the packing lemma inequality (\ref{eq:QpackB}),  %
the post-measurement state  $\sigma'^{\, m_0,\ell_1,\ell_2 }_{E_1^n}$ is close to the original state  $\sigma^{m_0,\ell_1,\ell_2}_{E_1^n}$, before the measurement $\Xi_{m_0}'$, in the sense that
\begin{align}
\frac{1}{2}\norm{\sigma'^{\, m_0,\ell_1,\ell_2 }_{E_1^n}-\sigma^{m_0,\ell_1,\ell_2}_{E_1^n}}_1 \leq 2^{ -n\frac{1}{2}( I(X_0;E_1)_\rho -R_1-\eps_3(\delta)) } \leq \eps_4(\delta)
\end{align}
for sufficiently large $n$ and rate as in (\ref{eq:E2}).
  Given that the event $\mathscr{G}_A^c$ has occurred, we have $(X_0^n(1),X_1^n(1,1))\in\Aset^{
 \nicefrac{\delta}{2}}(p_{X_0,X_1})$. %
 Now, by the basic properties of type class projectors, %
 \begin{align}
 \Pi^{\delta}(\rho_{E_1}|x_0^n)  \rho^{x_0^n}_{E_1^n}   \Pi^{\delta}(\rho_{E_1}|x_0^n) \preceq& 2^{ -n(H(E_1|X_0)_{\rho}-\eps_5(\delta)) } \Pi^{\delta}(\rho_{E_1}|x_0^n)
 \\
 \trace\left[ \Pi^{\delta}(\rho_{E_1}|x_0^n,x_1^n) \rho_{E_1^n}^{x_0^n,x_1^n} \right] \geq& 1-\eps_5(\delta) \\
 \trace\left[ \Pi^{\delta}(\rho_{E_j}|x_0^n,x_1^n)  \right] \leq& 2^{ n(H(E_j|X_0,X_1)_{\rho} +\eps_5(\delta))} \\
 \trace\left[ \Pi^{\delta}(\rho_{E_j}^{x_0^n}) \rho_{E_j^n}^{x_0^n,x_1^n} \right] \geq& 1-\eps_5(\delta) 
 \end{align}
 for all $(x_0^n,x_1^n)\in\Aset^{\delta_1}(p_{X_0,X_2})$, for $j=0,1$, (see %
 (\ref{eq:UnitTCond})-%
 (\ref{eq:UnitTCondB})). %
  By the quantum packing lemma, Lemma~\ref{lemm:Qpacking}, there exists a POVM $\Xi''_{\ell_1|\tm_0}$, for Eve 1, such that
 \begin{align}
 \cprob{ \mathscr{D}_{E_1,1} }{ \mathscr{A}^c\cap \mathscr{D}_{E_1,0}^c } \leq 2^{ -n( I(X_1;E_1|X_0)_\rho -\tR_1-\eps_6(\delta)) } \,.
 \end{align}
 This tends to zero as $n\rightarrow\infty$, provided that 
 \begin{align}
 \tR_1< I(X_1;E_1|X_0)_\rho -\eps_6(\delta) \,.
 \label{eq:E1}
 \end{align}
Based on the definition of subcodebooks, decoding $M_0$ and $L_1$ correctly guarantees that the layer-1 message $M_1$ will be decoded correctly as well.
As before, we observe that $I(X_1;E_1|X_0)_\rho\leq I(X_1;B|X_0)_\rho$ since the channel is degraded.
Thus, the packing lemma guarantees that  there exists a POVM  $\Gamma_{\ell_1|\hm_0}''$, for Bob, such that
 $%
 \cprob{ \mathscr{D}_{B,1} }{ \mathscr{G}_A^c\cap \mathscr{D}_{B,0}^c } %
 $ %
 tends to zero as 
 $n\rightarrow\infty$.

We now address the secrecy requirement for the layer-1 confidential message $M_1$. By the quantum covering lemma, Lemma~\ref{lemm:Qcovering},
\begin{align}
 \cprob{ \mathcal{S}_{E_2,1} }{ \mathscr{A}^c\cap\mathscr{D}_{E_2,0}^c  } \leq \exp\left(-2^{ n(\tR_1-R_1- I(X_1;E_2|X_0)_\sigma -\eps_7(\delta))}\right)  
 \end{align}
 which tends to zero as $n\rightarrow\infty$, provided that 
 \begin{align}
 R_1<\tR_1- I(X_1;E_2|X_0)_\rho -\eps_7(\delta) \,.
 \label{eq:E3R}
 \end{align}
The complementary event, $\mathcal{S}_{E_2,1}^c\equiv \{\norm{ \sigma_{E_2^n}^{\ell_1}-\breve{\sigma}_{E_2^n} }_1\leq \delta\}$, implies that the layer-1 leakage rate is bounded by $s^{(1)}(\Fset)\equiv I(M_1;E_2^n|M_0)_\rho\leq \eps_8(\delta)$ which tends to zero as $\delta\rightarrow 0$, due to the continuity of the quantum entropy (see Alicki-Fannes-Winter inequality \cite{AlickiFannes:04p,Winter:16p} \cite[Theorem 11.10.3]{Wilde:17b}).
By (\ref{eq:E1}) and (\ref{eq:E3R}), the layer-1 message $M_1$ can be transmitted reliably and in secret from Eve 2, provided that $R_1<I(X_1;E_1|X_0)_\rho - I(X_1;E_2|X_0)_\rho -\eps_6(\delta)-\eps_7(\delta)$.

Next, we consider the error event for the confidential message $M_2$, in the highest secrecy layer.
As with Eve 1, we use the the gentle measurement lemma %
to claim that
the post-measurement state  $\sigma'^{\,\ell_1,\ell_2 }_{B^n}$ is close to the original state  $\sigma^{\ell_1,\ell_2}_{B^n}$, before the measurement $\Gamma_{m_0}'$, \ie
\begin{align}
\frac{1}{2}\norm{\sigma'^{\,\ell_1,\ell_2 }_{B^n}-\sigma^{\ell_1,\ell_2}_{B^n}}_1 \leq 2^{ -n\frac{1}{2}( I(X_1;B|X_0)_\rho -\tR_1-\eps_6(\delta)) } \leq \eps_9(\delta)
\end{align}
for sufficiently large $n$ and rate as in (\ref{eq:E1}).
Once more, by the quantum packing lemma, there exists a POVM $\Gamma_{\ell_2|\hat{\ell}_1,\hm_0}'''$ such that
 \begin{align}
 \cprob{ \mathscr{D}_{B,2} }{ \mathscr{G}_A^c\cap \mathscr{D}_{B,0}^c\cap \mathscr{D}_{B,1}^c } \leq 2^{ -n( I(X_2;B_1|X_0,X_1)_\rho -\tR_2-\eps_{10}(\delta)) } 
 \label{eq:error2bound}
 \end{align}
  This tends to zero as $n\rightarrow\infty$, provided that 
 \begin{align}
 \tR_2&< I(X_2;B|X_0,X_1)_\rho -\eps_{10}(\delta) 
 \,.
 \label{eq:B}
 \end{align}

 As for the layer-1 secrecy requirement, by the quantum covering lemma, Lemma~\ref{lemm:Qcovering},
$%
 \cprob{ \mathcal{S}_{E_1 E_2,2} }{ \mathscr{D}_{E_1,0}^c\cap\mathscr{D}_{E_2,0}^c\cap \mathscr{D}_{E_1,1}^c } \leq \exp \left(- 2^{ n(\tR_2-R_2- I(X_2;E_1 E_2|X_0,X_1)_\sigma -\eps_{11}(\delta)) } \right) 
 $, %
 which tends to zero as $n\rightarrow\infty$, provided that 
 \begin{align}
 R_2&<\tR_2- I(X_2;E_1 E_2|X_0,X_1)_\rho -\eps_{11}(\delta) 
 \label{eq:E3R2}
 \end{align}
 By (\ref{eq:B})-(\ref{eq:E3R2}), it suffices that 
 $R_2<I(X_2;B|X_0,X_1)_\rho- I(X_2;E_1 E_2|X_0,X_1)_\rho -\eps_{10}(\delta)  -\eps_{11}(\delta) $.
 The achievability proof is completed by taking $n\rightarrow\infty$ and then $\delta\rightarrow 0$.
 
 \subsection{Converse proof }
  Suppose that Alice chooses layer-0, 1, and 2 messages, $M_0$, $M_1$, and $M_2$, uniformly at random. She prepares an input state $\rho^{m_0,m_1,m_2}_{ A^n}$. 
The channel output is
$\rho_{B^n E_1^n E_2^n}=\frac{1}{2^{n(R_0+R_1+R_2)}}\sum_{m_0,m_1,m_2} \channel_{A^n\rightarrow B^n E_1^n E_2^n}(\rho^{m_0,m_1,m_2}_{ A^n}) $.
Then, Bob, Eve 1, and Eve 2 perform  decoding POVMs $\Gamma^{m_0,m_1,m_2}_{B^n}$,  $\Xi^{m_0,m_1}_{E_1^n}$, and $\Upsilon^{m_0}_{E_2^n}$, respectively. As a measurement outcome, Bob obtains his estimation $(\hM_0,\hM_1,\hM_2)$, Eve 1 procures $(\tM_0,\tM_1)$, and Eve 2 measures $\breve{M}_0$.

Consider a sequence of layered-secrecy codes $(\Fset_n,\Gamma_n,\Xi_n,\Upsilon_n)$, such that the average probability of error and the leakage rates tend to zero, hence
the error probabilities $\prob{ \breve{M}_0\neq M_0}$, $\prob{ \tM_1\neq M_1 }$,  $\prob{ \hM_2\neq M_2 }$,     are bounded by some
$\alpha_n$ which tends to zero as $n\rightarrow \infty$.
By Fano's inequality \cite{CoverThomas:06b}, it follows that
\begin{align}
H(M_0|\breve{M}_0) \leq n\eps_{1n}
\label{eq:Hm0ukEve2LS}
\\
H(M_1|\tM_1) \leq n\eps_{2n}
\label{eq:Hm0ukEve1LS}
\\
H(M_2|\hM_2) \leq n\eps_{3n}
\label{eq:Hm1ukLS}
\end{align}
where $\eps_{jn}$ tend to zero as $n\rightarrow\infty$.
Furthermore, the leakage rates are bounded by
\begin{align}
    I(M_1;E_2^n|M_0)_\rho\leq \delta_n
    \label{eq:leakConU1kLSe2}\\
    I(M_2;E_1^n E_2^n|M_0)_\rho\leq \delta_n
    \label{eq:leakConU1kLSe1e2}
\end{align}
where $\delta_n$ tends to zero as $n\rightarrow \infty$.

Thus, the layer-0 common rate is bounded as 
\begin{align}
nR_0&=H(M_0)
\nonumber\\
&=I(M_0;\breve{M_0})+H(M_0|\breve{M_0})\nonumber\\
&\leq I(M_0;E_2^n)_{\rho}+n\eps_{1n} 
\label{eq:R0convEkLS}
\end{align}
where the first equality holds since the messages are uniformly distributed by assumption.
In the last line, we bounded the first term by $I(M_0;E_2^n)_{\rho}$, using the data processing inequality, and the second term by $\eps_{1n}$ using (\ref{eq:Hm0ukEve2LS}).

Similarly, the layer-1 confidential rate is bounded as
\begin{align}
    nR_1 
    &=H(M_1|M_0)
    \nonumber\\
    &=I(M_1;\tM_1|M_0)+H(M_1|\tM_1,M_0)
    \nonumber\\
    &\leq I(M_1;\tM_1|M_0)+H(M_1|\tM_1)
    \nonumber\\
    &\leq I(M_1;E_1^n|M_0)_\rho+n\eps_{2n}
    \label{eq:nR1kU1LS} 
\end{align}
where the first equality follows from  the  statistical independence between the messages, and the first inequality holds since 
conditioning cannot increase entropy \cite[Corollary 11.8.1]{Wilde:17b}. 
As seen in (\ref{eq:leakConU1kLSe2}),  $-I(M_1;E_2^n|M_0)_\rho+ n\delta_n\geq 0$ due to the layer-1 secrecy requirement. Hence, we deduce that 
\begin{align} 
nR_1&\leq I(M_1;E_1^n|M_0)_\rho-I(M_1;E_2^n|M_0)_\rho+n\delta_n+n\eps_{2n} 
\label{eq:nR1upp0kLS}
\end{align}

By the same considerations, the top-secret layer-2 confidential rate is bounded as
\begin{align}
    nR_2 
    &\leq I(M_2;\hM_2|M_0,M_1)+H(M_2|\hM_2)
    \nonumber\\
    &\leq I(M_2;B^n|M_0,M_1)_\rho+n\eps_{3n}
    \nonumber\\
    &\leq I(M_2;B^n|M_0,M_1)_\rho-I(M_2;E_1^n|M_0,M_1)_\rho+n\delta_n+n\eps_{3n}
    \label{eq:nR2kU1LS} 
\end{align}
where the last inequality follows from the layer-2 secrecy requirement in (\ref{eq:leakConU1kLSe1e2}).
This completes the proof of Theorem~\ref{theo:LS}.
\qed

\end{appendices} 

\bibliography{references2}{}

\end{document}

%% file: preambles.tex
\usepackage{amsthm}
\usepackage{amsmath}
\usepackage{mathrsfs}
\usepackage{amssymb} 
\usepackage{bbm,dsfont} 
\usepackage{multirow} 
\usepackage{units}
\usepackage{stmaryrd}   
\usepackage{verbatim}
\usepackage{enumerate} 
\usepackage{xcolor}
\usepackage{tikz}
\usetikzlibrary{calc}
\usepackage{tabularx}
\usepackage{ifthen}
\usepackage{accents}

\usepackage{balance} 

\usepackage[sort&compress,square,numbers]{natbib}
\usepackage{graphicx}
\graphicspath{{images/}{figures/}{./}{./ITW_figures}}

\newcommand{\prob}[1]{\Pr\left(#1\right)}
\newcommand{\given}{\mid}
\newcommand{\cprob}[2]{\Pr\left(#1\given #2\right)}

\renewcommand{\mathbbm}[1]{\text{\usefont{U}{bbm}{m}{n}#1}} 
\newcommand{\eps}{\varepsilon}

\newcommand{\norm}[1]{\left\lVert#1\right\rVert}
\newcommand{\trace}{\mathrm{Tr}}

\newcommand{\identity}{\mathbbm{1}}
\newcommand{\ket}[1]{ | #1 \rangle }
\newcommand{\bra}[1]{ \langle #1 | }
\newcommand{\ketbra}[1]{ | #1 \rangle\langle #1 | } 

\newcommand{\ie}{\emph{i.e.}, }
\newcommand{\eg}{\emph{e.g.}, }
\newcommand{\etal}{\emph{et al.} }

\newcommand{\tk}{\widetilde{k}}	
\newcommand{\tm}{\widetilde{m}}	
\newcommand{\tM}{\widetilde{M}}
\newcommand{\tK}{\widetilde{K}}

\newcommand{\tX}{\widetilde{X}}

\newcommand{\tR}{\widetilde{R}}

\newcommand{\ha}{\hat{a}}
\newcommand{\hb}{\hat{b}}
\newcommand{\hc}{\hat{c}}
\newcommand{\hd}{\hat{d}}
\newcommand{\he}{\hat{e}}

\newcommand{\hk}{\hat{k}}
\newcommand{\hm}{\hat{m}}
\newcommand{\hn}{\hat{n}}
\newcommand{\hs}{\hat{s}}

\newcommand{\hK}{\hat{K}}
\newcommand{\hP}{\hat{P}}
\newcommand{\hM}{\hat{M}}
\newcommand{\hL}{\hat{L}}

\newcommand{\Aset}{\mathcal{A}}
\newcommand{\Bset}{\mathcal{B}}

\newcommand{\Dset}{\mathcal{D}}
\newcommand{\Fset}{\mathcal{F}}

\newcommand{\Gset}{\mathcal{G}}
\newcommand{\Hset}{\mathcal{H}}

\newcommand{\Kset}{\mathcal{K}}

\newcommand{\Uset}{\mathcal{U}}

\newcommand{\Tset}{\mathcal{T}}	

\newcommand{\Xset}{\mathcal{X}}
\newcommand{\Yset}{\mathcal{Y}}
\newcommand{\Zset}{\mathcal{Z}}

\newcommand{\markovC}[1]{%
\begin{tikzpicture}[#1]%
\draw (0,0.3ex) -- (1ex,0.3ex);%
\draw (0.5ex,0.3ex) circle (0.2ex);
\draw[white] (0.2ex,0) -- (0.5ex,0);%
\end{tikzpicture}%
}
\newcommand{\Cbar}{\markovC{scale=2}}

\newcommand{\tset}{\Aset^{\delta}}										

\newlength{\dhatheight}

\theoremstyle{remark}	\newtheorem{theorem}{Theorem}
\theoremstyle{remark}	\newtheorem{lemma}[theorem]{Lemma}
\theoremstyle{remark}	\newtheorem{corollary}[theorem]{Corollary}
\theoremstyle{remark}	
\theoremstyle{remark}	\newtheorem{conjecture}{Conjecture}
\theoremstyle{remark} \newtheorem{definition}{Definition}
\theoremstyle{remark} \newtheorem{remark}{Remark}
\theoremstyle{remark}

\newcommand{\channel}{\mathcal{L}}

\newcommand{\inC}{\mathsf{C}}	
\newcommand{\inR}{\mathsf{R}}	
\newcommand{\opC}{\mathcal{C}} 

